\documentclass[10pt,a4paper,twocolumn ]{article}
\usepackage[T1]{fontenc}
\usepackage{inputenc}
\usepackage[english]{babel}
\usepackage{amsmath}
\usepackage[nohead,includefoot,margin=2cm]{geometry}
\usepackage[compact,raggedright,small]{titlesec}
\usepackage[affil-it]{authblk}
\usepackage[runin]{abstract}
\usepackage{multicol}
\usepackage{blindtext}

\addto{\Affilfont}{\small}

\usepackage{txfonts}

\usepackage{graphicx}
\usepackage{amssymb}
\usepackage{bookmark}
\usepackage{bm}
\usepackage{hyperref}
\usepackage{algorithm, algpseudocode}
\usepackage{subcaption}
\usepackage{url}
\usepackage{csvsimple}
\usepackage{stfloats}
\usepackage{tcolorbox}
\usepackage{overpic}
\usepackage{xcolor, colortbl}
\titleformat*{\paragraph}{\itshape\mdseries}

\tcbset{colback=white,colframe=black}

\definecolor{lightlightgray}{rgb}{0.9, 0.9, 0.9}

\title{Automatic cable harness layout routing in a customizable 3D environment}
\author[1]{T. Karlsson}
\author[1]{E. Åblad}
\author[1]{T. Hermansson}
\author[1]{J. S. Carlson}
\author[2]{G. Tenfält}
\affil[1]{Fraunhofer-Chalmers Research Centre for Industrial Mathematics, Götebotg, 412 88,Sweden}
\affil[2]{Volvo Car Corporation, Göteborg, 405 31, Sweden}
\date{\small \today}

\begin{document}
\twocolumn[
	\begin{@twocolumnfalse}
	\maketitle
\begin{abstract}
Designing cable harnesses can be time-consuming and complex due to many design and manufacturing aspects and rules. Automating the design process can help to fulfil these rules, speed up the process, and optimize the design. To accommodate this, we formulate a harness routing optimization problem to minimize cable lengths, maximize bundling by rewarding shared paths, and optimize the cables' spatial location with respect to case-specific information of the routing environment, e.g., zones to avoid. A deterministic and computationally effective cable harness routing algorithm has been developed to solve the routing problem and is used to generate a set of cable harness topology candidates and approximate the Pareto front. Our approach was tested against a stochastic and an exact solver and our routing algorithm generated objective function values better than the stochastic approach and close to the exact solver. Our algorithm was able to find solutions, some of them being proven to be near-optimal, for three industrial-sized 3D cases within reasonable time (in magnitude of seconds to minutes) and the computation times were comparable to those of the stochastic approach.

\end{abstract}
 \end{@twocolumnfalse}
]

\section{Introduction} \label{sec:introduction}

\subsection{Industrial background} \label{secsub:industrial_background}
\begin{figure}[]
	\centering
	\begin{subfigure}[t]{0.5\textwidth}
		\centering
		\includegraphics[width=\linewidth]{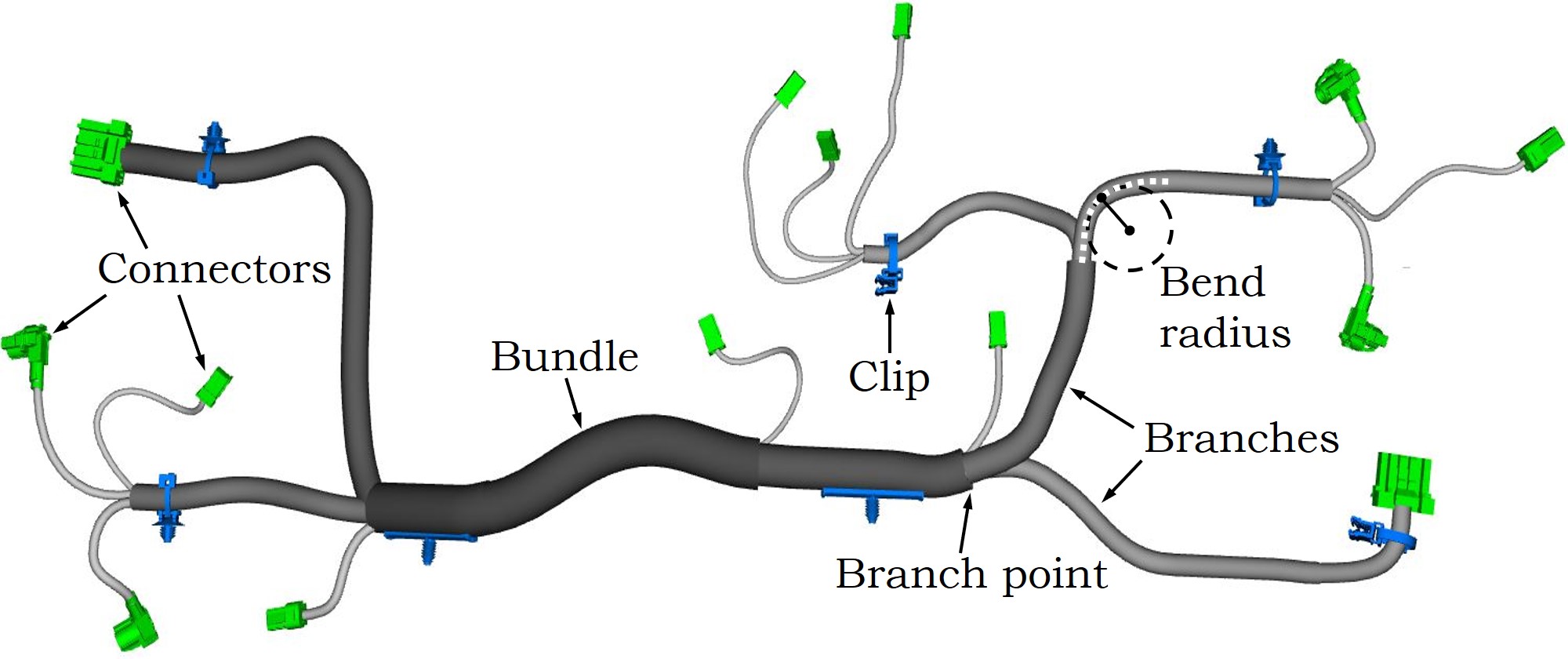}
	\end{subfigure}
	\caption{A cable harness and its terminology.}
	\label{fig:harness_components}
\end{figure}

\begin{figure}[]
	\centering
	\begin{subfigure}[t]{0.4\textwidth}
		\centering
		\includegraphics[width=\linewidth]{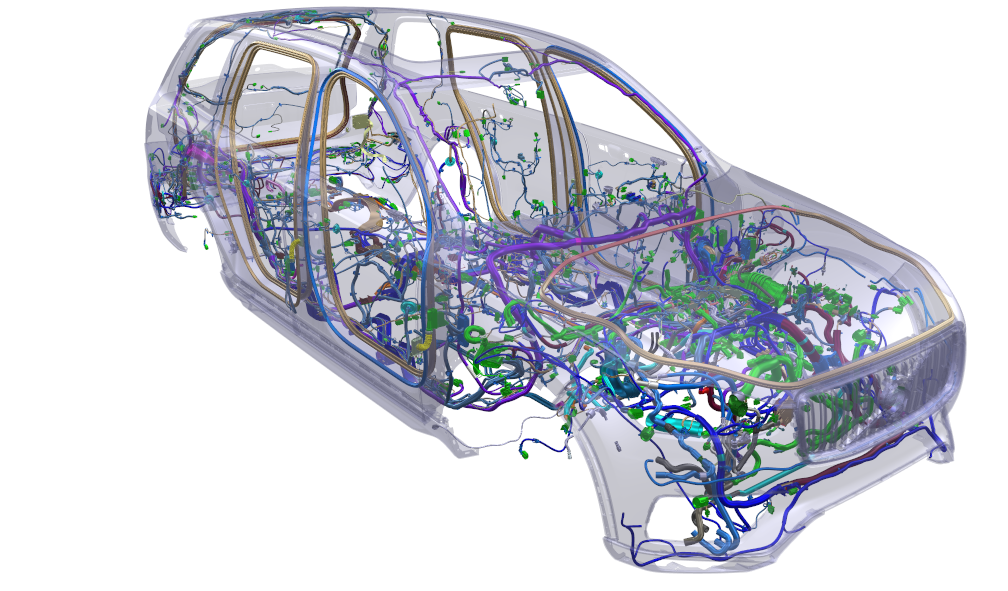}
	\end{subfigure}
	\caption{Cables and wire harnesses in a Volvo XC90.}
	\label{fig:car_harness}
\end{figure}

Automotive, aerospace, telecommunications, medical, and robotics are some demanding applications for mechanical and electrical design. These applications contain systems of interconnected elements including electronic components, control units, sensors, and actuators. Structured assemblies of cables and wires bundled together, called cable harnesses, are designed to simplify these connections, while at the same time protecting the wires. Designing cable harnesses is a repetitive and time-demanding process and can be complex due to many design and manufacturing aspects and rules~\cite{ng2000designing, pemarathne2016wire, zhu2017methodology}. Harness layouts are designed with respect to the length of the cables, harness weight, bundling, bend radii, the position of fasteners (clips or clamps), avoiding high-temperature zones, electrical requirements and performance (e.g., crosstalk and voltage drop), and material cost, among others~\cite{ng2000designing, pemarathne2016wire, zhu2017methodology, pradhan2011current}. A cable harness is shown in Figure~\ref{fig:harness_components}. The design complexity increases when a higher number of cables need to be routed in the same design space, this is an issue in, e.g., the automotive industry where the number of electrical components and wires has steadily risen~\cite{pradhan2011current, braun2015trends, trommnau2019overview}. A harness routing from the automotive industry is shown in Figure~\ref{fig:car_harness}. The routing problem is further complicated due to influences from decisions made regarding a product prototype. Minor changes that affect, for example, a chassis design or an individual component can require a new harness configuration~\cite{ng2000designing}. Harness routing issues can even result in late and expensive re-design of a chassis~\cite{ng2000designing}. To avoid late detection of assembly installation problems, both feasible installation paths and ergonomic aspects of these paths must be considered in the routing phase~\cite{hermansson2013automatic}. The wire harness manufacturing process (i.e., the assembling of wires and other materials into a harness) is partly automated in the industry~\cite{NGUYEN2021379} and the harness design influences the achievable degree of manufacturing automation~\cite{trommnau2019overview}.

The cable harness layout routing problem involves determining bundle paths, i.e., routes, and branch point locations of a harness within a defined geometric space to connect various components, all while considering the numerous design aspects and rules mentioned above. Automating the layout design process can help to reduce the lead time and fulfil all design constraints. This paper focuses on the automated design of the so-called harness topology, in which the main issue is the trade-off between individual cable lengths and common lengths (bundling). The topology is determined while taking into account design space requirements such as clip-able surfaces and installation-friendly zones. Not all design aspects and rules are considered at this stage, hence several topology solutions are suggested and serve as the initial step in the design process for further processing.

\subsection{Related work}
Masoudi and Fadel~\cite{masoudi2022optimization} classified the state-of-the-art studies related to cable harness design into two main classes (that are not mutually exclusive), namely design process and layout optimization. The design process includes tools that deal with the testing of prototypes in virtual and augmented reality environments~\cite{ng2000designing, valentini2011interactive, ritchie2007cableCostar, yang2019review} and knowledge-based or artificial intelligence design approaches~\cite{pemarathne2016wire, shang2012computational, park_cutkosky_conru_lee_1994, van2007intelligent} to support designers to satisfy design rules and make design-related decisions.

This paper focuses on cable harness layout optimization, where a mathematical model is formulated that includes several important harness design factors. Examples of harness factors that have been modelled as objective functions or constraints in previous work include the total length and weight of the harness, shared path lengths (bundling), installation aspect (by avoiding routing in narrow spaces), electrical factors (e.g., power, resistance, and crosstalk), clipping rules, minimum bend radius, and the cost of materials (e.g., cables, clips, and protection layers)~\cite{conru1994genetic, zhang2021multi, zhang2021multi_PS,  masoudi2022optimization, lin2015efficient, jin2021cable, zhu2017methodology, kim2021sequential, castorani2018method, eheim2021automation, dinkelacker2023system, fedorov2017vehicle}. Given a layout routing, methods have been proposed to optimize cable gauges and cross-section of bundles to minimize harness weight while satisfying requirements regarding currents, voltage drop, and temperature~\cite{zozaya1994use, combettes2015multi, rius2016optimization}. Conru and Cutkosky~\cite{conru_computational} discussed the problem of case-specific constraints and objectives not encoded in their router, and how human input can be used to guide the search towards a desired layout. Zhu et al.~\cite{zhu2017methodology} accounted for various aspects of the routing environment, such as high-temperature zones where protection covers for the harness are required, resulting in an additional cost. Maier and Reuss~\cite{maier2023design} used the edge costs in their graph search problem to represent requirements such as predefined main cable paths.

The harness routing problem can be seen as finding a Steiner tree, where a Steiner point represents a branch point~\cite{zhang2021multi, zhang2021multi_PS}. Steiner tree approaches have also been used for branch pipe routing~\cite{liu2012multi, blokland2023literature} and electrical wire routing~\cite{lin2015efficient, huang2016delay}. Conru~\cite{conru1994genetic} notes that although the harness routing problem has similarities to the \emph{Steiner tree problem (STP)}, they sufficiently differ to inhibit the direct use of search methods applied to Steiner trees since the cost associated with an edge in the harness routing graph depends on the number of wires in the traversing bundle (which is not known a priori). \emph{Particle Swarm Optimization (PSO)} has been used to search for Steiner trees~\cite{zhang2021multi, zhang2021multi_PS, liu2012multi}. A PSO solver has been implemented in this paper for comparison and this approach is explained in detail in \ref{appendix:pso}. Other stochastic approaches that have been proposed for cable harness routing include evolutionary and genetic algorithms~\cite{weise2019graph, conru1994genetic, zhao2021multi, masoudi2022optimization, jiang2023branch} and an ant colony optimization algorithm~\cite{komninou2011optimal}. Ant colony algorithms have also been used for multi-hose routing to minimize total length and maximize shared lengths~\cite{fernando2012multi}. Regarding deterministic approaches, Lin et al.~\cite{lin2015efficient} solved a wire routing problem formulated as an STP where a Steiner point represents the location of a splice that is used to connect more than two wires; they used a modified Kou-Markowsky-Berman algorithm and a \emph{Linear Programming (LP)} solver to search for Steiner trees. Zhu et al.~\cite{zhu2017methodology} used a local search algorithm called hill climbing to optimize branch point locations, an $A^*$ algorithm to optimize branch paths, and pattern search to further optimize the harness routing and to eliminate violations of some design rules. Parque et al.~\cite{parque2019path} used the deterministic blackbox algorithm DIRECT, along with various stochastic methods, to address the so-called path bundling problem within polygonal domains. The key motivation for their algorithms was to handle minimal-length tree problems such as wire harness routing. Rehal and Sen~\cite{rehal2023generation} presented a topological method to find all non-homotopic paths between multiple pairs of starting and ending points---this enables them to generate bundles through continuous deformation. However, they left the task of automatically placing branch points for future work.

This paper uses Lagrangian relaxation together with a subgradient method to optimize the Lagrangian dual problem. A Lagrangian relaxation of an integer programming problem produces a Lagrangian (sub)problem (i.e., computing the dual function value given a dual solution) that can be solved and whose value is a lower bound on the optimal value of the original problem~\cite{fisher1981lagrangian}. The dual function is concave (assuming the primal problem is to minimize) but non-smooth. Maximizing the dual function, and hence finding the best lower bound, is known as the dual problem and can be done by using the subgradient method. This methodology has been applied to combinatorial optimization problems such as the generalized travelling salesperson problem~\cite{noon1991lagrangian} and the STP~\cite{moshe1990Relax}. The reasons for using these techniques in this paper are to observe the lower bound on the optimal primal objective function value and to use the Lagrangian subproblem solutions to find topologically different primal solutions.

A multi-objective optimization problem is formulated in this paper, and various techniques are available for handling multiple objective functions and exploring the Pareto front~\cite{chankong_multiobjective, gunantara2018review}. We use the weighted sum method and there exist methods to find solutions well-distributed along the Pareto front~\cite{kim2005adaptive, cai2018grid}. However, for simplicity and sufficiency within the scope of this paper, we optimize the corresponding scalar problem using an equally distributed set of weights.

\subsection{Contribution, scope, and outline} \label{sec:scope}
The paper includes objective functions regarding cable lengths and bundling, along with the encoding of routing environmental information, similar to previous work. The novelty of our approach is the way we model these aspects, allowing us to utilize our developed efficient algorithms. In alignment with previous work on the STP, which is a specific instance of our model, we utilize a Lagrangian relaxation, thus generalizing their approach. It is mainly investigated how the developed algorithms perform with respect to the computation time and the objective function value, and how we can generate topologically distinct solutions. The paper discusses how certain design factors can be considered through soft constraints, such as clip-able and cable-safe zones. We define a harness topology as a collection of cables and their routes, resulting in a number of branch points.

The paper has the following outline: Section \ref{sec:method} describes the mathematical optimization problem and the design aspects that are included. Algorithms to find solutions to the problem are presented in Section \ref{sec:algorithm} and their performances are analysed in Section \ref{sec:results}. Section \ref{sec:results} also presents further design aspects and rules, and how they can be handled in a design framework. In Section \ref{sec:conclusion_and_future_work}, we conclude the main findings and discuss future work. 

\section{Mathematical model} \label{sec:method}
The model presented in Section \ref{sec:model} and the algorithms in Section \ref{sec:algorithm} are based on previous work from the authors in~\cite{karlsson2020optimization}.

\subsection{The routing environment, grid graph, and cost field}
The 3D routing environment consists of pairs of start and end points (also known as terminals) between which the cables are to be routed, physical obstacles, and customizable environmental information or preferences. The terminals represent the locations of connectors. The environmental information may include cable-safe zones, ergonomic installation zones, and clip-able surfaces (where the harness is allowed to be fastened). The routing environment is discretized, within a bounding box enclosing the terminals, and represented using a graph. While any graph structure can be used, for simplicity, we create a regular grid with cubic cells, which we then scale to obtain a Cartesian grid. The grid graph has node set $V\subset\mathbb{N}^3$ corresponding to the discrete grid positions and an edge set $E$ defined as
\begin{align*}
	& E := \{ \{u,v\} \mid \|u-v\|_\infty = 1,\: u,v\in V\}.
\end{align*}
The environmental information is mapped to a so-called cost field that corresponds to the node costs. The edge costs $c_e\in \mathbb{R}_+,\: e\in E$, depend on the node costs and physical Euclidean lengths. Nodes and edges intersecting with physical obstacles are removed from the graph, ensuring that the routes are collision-free.

\begin{figure}[!ht]
	\centering
	\begin{subfigure}[b]{0.5\textwidth}
		\centering
		\includegraphics[width=\textwidth]{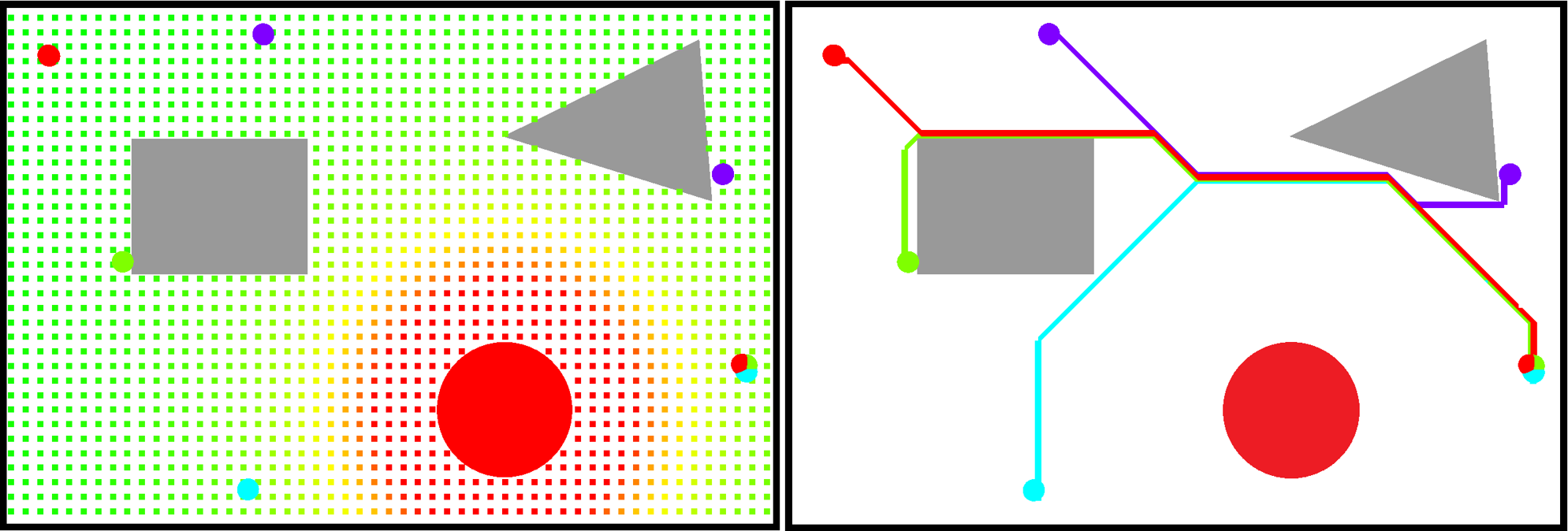}
		\caption{Avoid a high-temperature object (red) and zone, and do not collide with the grey objects.}
	\end{subfigure}
	\begin{subfigure}[b]{0.5\textwidth}
		\centering
		\includegraphics[width=\textwidth]{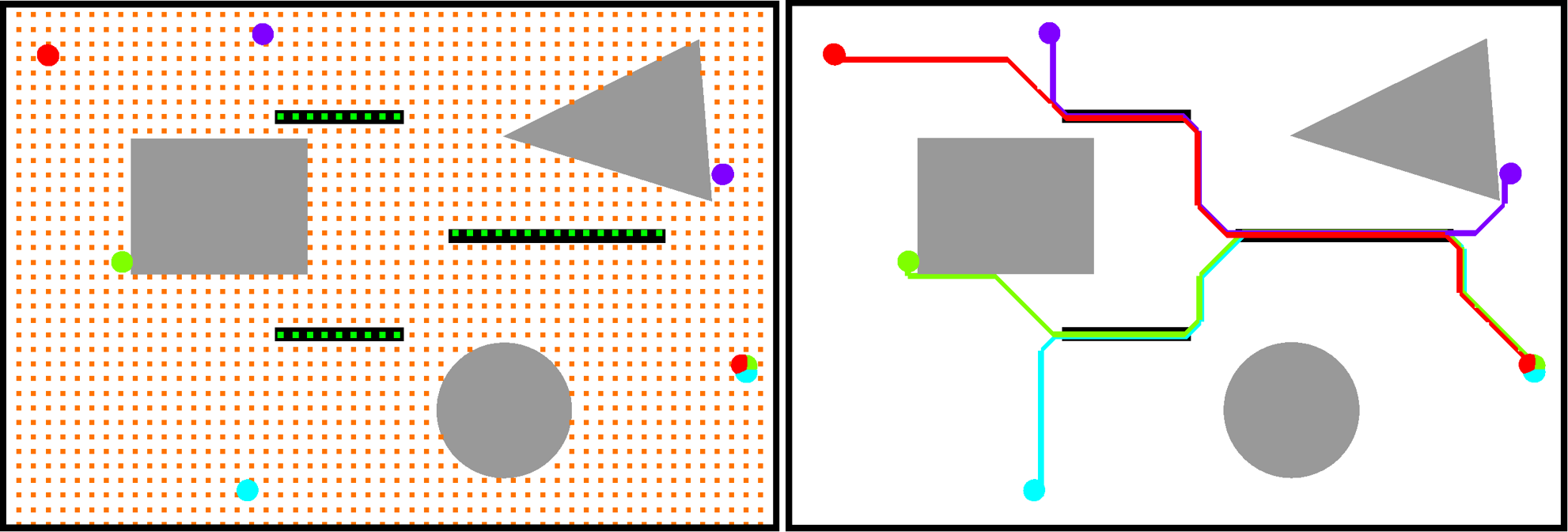}
		\caption{Prefer to route in cable trays (black).}
	\end{subfigure}
	\begin{subfigure}[b]{0.5\textwidth}
		\centering
		\includegraphics[width=\textwidth,height=2.5cm]{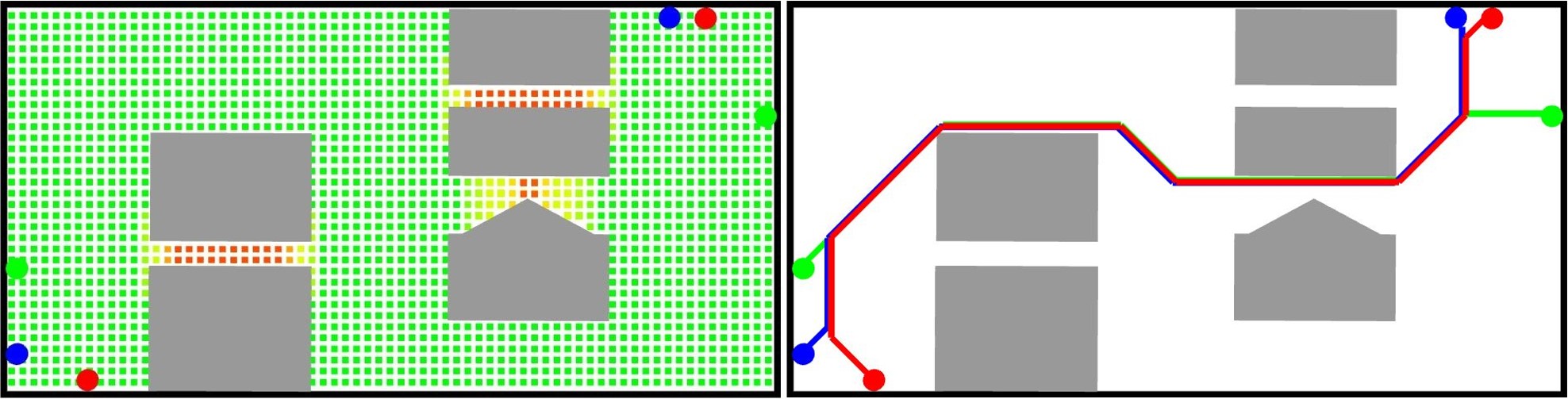}
		\caption{Prefer open spaces and avoid narrow areas for easier installation.}
	\end{subfigure}
	\caption{Customized cost fields and optimized routings between pairwise start and end nodes. Green and red grid points correspond to low and high costs, respectively.}
	\label{fig:different_costs}
\end{figure}

Figure~\ref{fig:different_costs} depicts three scenarios with customized environmental information mapped to suitable cost fields where low (green) cost and high (red) cost represent preferred and non-preferred zones to route in, respectively. Distances between the grid points and obstacles are mapped to positive real values that depend on the layout of the surrounding obstacles and their classification (e.g., hot objects and cable trays). An edge cost is calculated as the product of the Euclidean distance between the two connected nodes and the mean of their corresponding node costs. This approach allows us to minimize the physical length of the cables while considering routing environment requirements.

\subsection{Bi-objective cable harness routing optimization model} \label{sec:model}
The optimization problem is formulated as a bi-objective mixed-binary linear programming model on a graph. The problem is to find paths between start and end node pairs that minimize the length of the cables ($f_L$) and the harness bundles ($f_B$)---these two conflicting aspects are formulated as the following objective functions

\begin{subequations} \label{eq:obj_functions}
\begin{alignat}{3}
	f_L(\bm{y}) & := \sum\limits_{k\in K}\sum\limits_{e\in E} c_{e}y_e^k,\enskip \text{and} \label{eq:obj_L} \\
	f_B(\bm{x}) & := \sum\limits_{e\in E}c_{e}x_e, \label{eq:obj_B}
\end{alignat}	
\end{subequations}

with decision variables $y_e^k\in[0,1]$ and $x_e\in\{0,1\}$, $e\in E,\: k\in K$, where $K$ represents the set of cables. The variable $y_e^k$ is equal to 1 if edge $e\in E$ is traversed by cable $k\in K$ and 0 otherwise. $\bm{y}$ are continuous due to the integrality property of model \eqref{eq:CHRP} concerning $\bm{y}$~\cite{karlsson2020optimization}. In other words, dropping integrality conditions on $\bm{y}$ does not alter the optimal solution. The variable $x_e$ is equal to 1 (selected) if edge $e\in E$ is traversed by at least one cable and 0 otherwise.

\begin{figure}[!ht]
	\centering
	\begin{subfigure}[b]{0.235\textwidth}
		\centering
		\begin{overpic}[width=\textwidth]{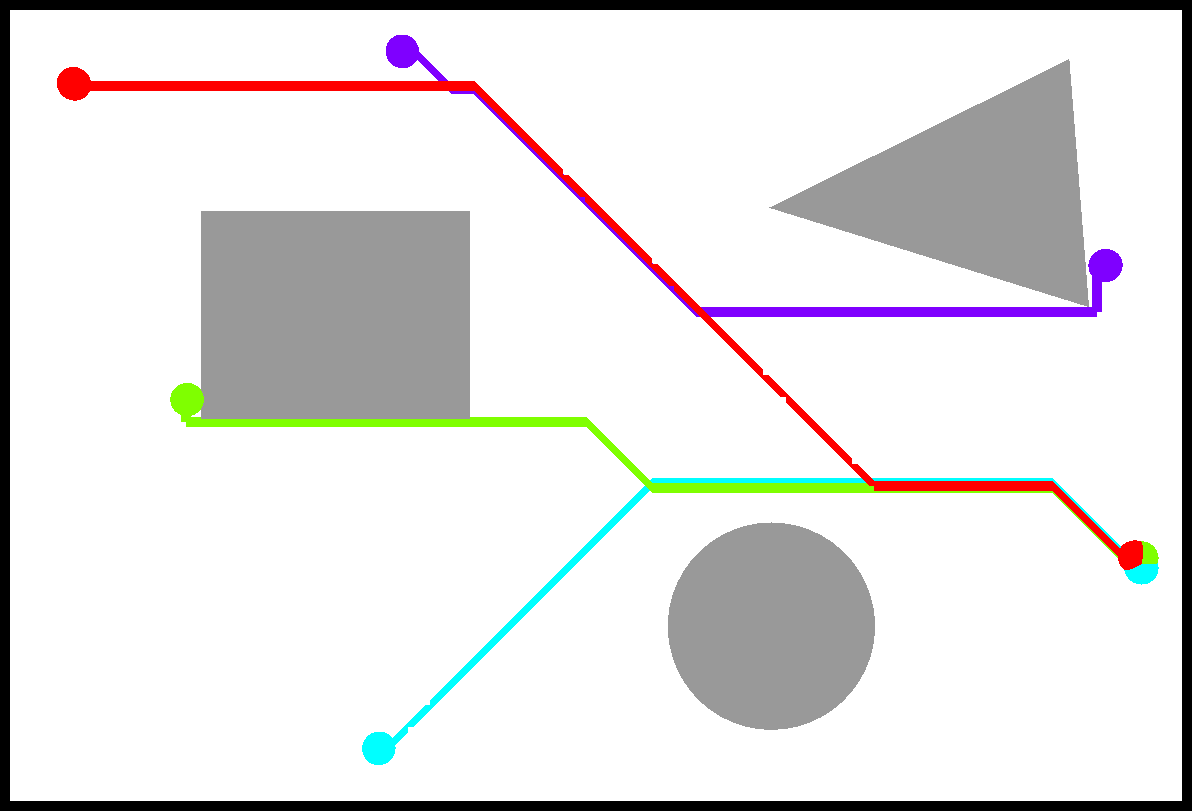}
			\put(44,59){Low $w_B$}
		\end{overpic}
		\caption{}
		\label{fig:different_weights_low_bundle}
	\end{subfigure} 
	\begin{subfigure}[b]{0.235\textwidth}
		\centering
		\begin{overpic}[width=\textwidth]{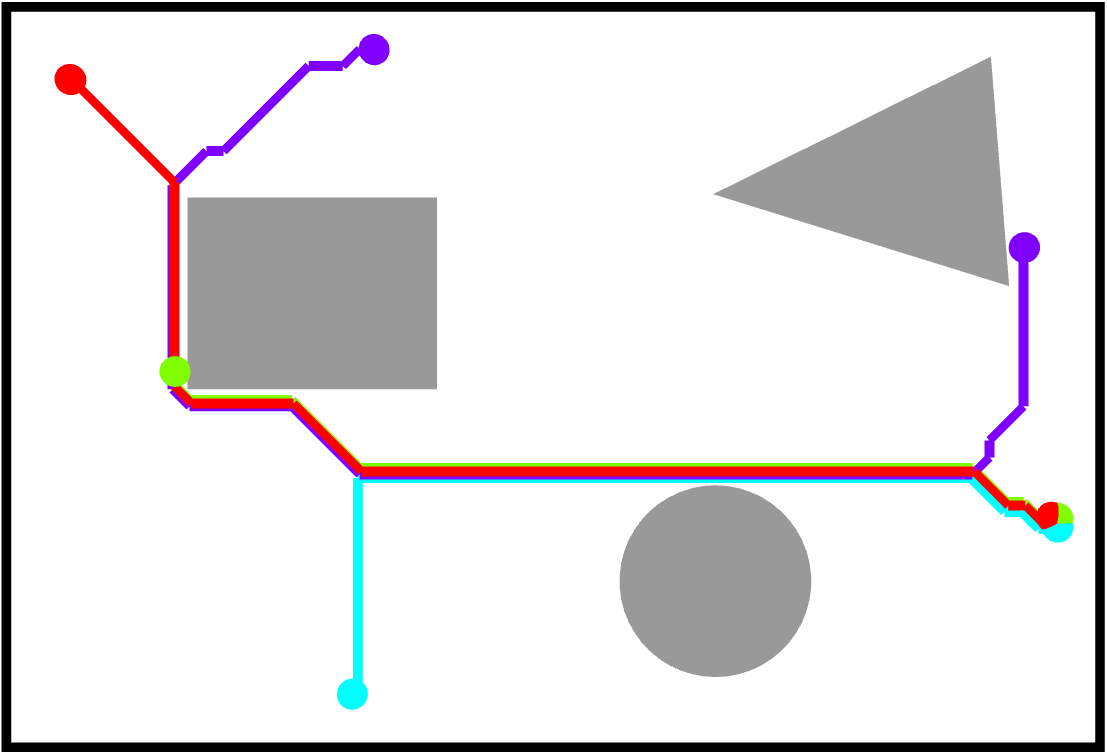}
			\put(44,59){High $w_B$}
		\end{overpic}
		\caption{}
		\label{fig:different_weights_high_bundle}
	\end{subfigure}
	\caption{Two optimized harness routings with (\subref{fig:different_weights_low_bundle}) low bundle weight and (\subref{fig:different_weights_high_bundle}) high bundle weight.}
	\label{fig:different_weights}
\end{figure}

The two objective functions are formulated as a single weighted function
\begin{equation} \label{eq:weighted_obj_fun}
	f(\bm{x},\bm{y}) := w_L f_L(\bm{y}) + w_B f_B(\bm{x}),
\end{equation}

with objective function weights $w_L,w_B\in \mathbb{R}_+$, where $w_L+w_B=1$. This is a common approach in multi-objective optimization to obtain a scalar problem, however, some Pareto optimal solutions may never be discovered by this procedure since the feasible set \eqref{eq:CHRP_con_paths}--\eqref{eq:CHRP_con_edgebin} is non-convex~\cite{chankong_multiobjective}.  Nevertheless, we can efficiently obtain topologically distinct solutions by applying the heuristic in Section \ref{sec:harness_heuristic} for a set of weights, as depicted in Figure~\ref{fig:different_weights} and \ref{fig:case_A_example_solution}.

The \emph{Cable Harness Routing Problem} (\emph{CHRP}) can be formulated as
\begin{subequations} \label{eq:CHRP}
	\begin{alignat}{3}
		\min			&& \enspace f(\bm{x}, \bm{y}), \label{eq:CHRP_obj} \\ 
		\textrm{s.t.} 	&& \bm{y}^k & \in 	 \mathcal{P}_k,						&& \quad k\in K, \label{eq:CHRP_con_paths} \\
		&&  y^k_e	& \leq 	 x_e,					 	&& \quad k\in K, \thinspace e \in E, \label{eq:CHRP_con_edgeact} \\
		&& \bm{x} 	& \in 	 \mathbb{B}^{\vert E\vert}.  \label{eq:CHRP_con_edgebin}
	\end{alignat}
\end{subequations}
With \eqref{eq:obj_L} we minimize the total cable lengths and with \eqref{eq:obj_B} we optimize the bundling. The usage or selection of an edge by at least one cable is penalized through \eqref{eq:obj_B}, and we prefer more bundling by increasing $w_B$. The value of the so-called bundle weight, $w_B$, can be seen as how much we prefer the cables to deviate from their shortest paths in favour of sharing common paths in a way that uses few and cheap edges, see Figure~\ref{fig:different_weights}. The set $\mathcal{P}_k$ represents the possible paths for cable $k\in K$ between its terminals (formulated as flow conservation constraints as in~\cite{karlsson2020optimization}). The so-called edge selection constraints \eqref{eq:CHRP_con_edgeact} ensure that the binary variable $x_e$ is selected if edge $e\in E$ is traversed by any cable. 

Note that if $w_B = 0$, the CHRP reduces to $\vert K \vert$ shortest path problems. If $w_B=1$, then \eqref{eq:weighted_obj_fun} is equivalent to the objective function for the STP, and the CHRP generalizes the multi-commodity flow formulation of the STP~\cite{wong1984dual} since not all end nodes need to have the same start, i.e., the solutions do not need to form a tree. The STP is NP-complete~\cite{moshe1990Relax}. Consequently, the CHRP inherits the NP-hard property, making it suitable for the use of heuristic approaches.

\subsubsection{Lagrangian relaxation of the CHRP}
Lagrangian relaxation is a technique that can be applied for large-scale integer programs to simplify the problem and obtain useful information but does not necessarily generate feasible or satisfactory primal solutions. The complicating edge selection constraints \eqref{eq:CHRP_con_edgeact} are relaxed and penalized with Lagrangian dual variables $\bm{\lambda}\geq\bm{0}$, resulting in the Lagrangian subproblem

\begin{equation} \label{eq:Lagrange_relax}
	\min \enspace \big\{ f(\bm{x}, \bm{y}) + \sum_{k\in K}\sum_{e\in E}\lambda_e^k (y_e^k - x_e) \mid \bm{x}  \in 	 \mathbb{B}^{\vert E\vert}, \: \bm{y}^k \in \mathcal{P}_k \big\}.
\end{equation}

As shown in~\cite{karlsson2020optimization,moshe1990Relax}, problem \eqref{eq:Lagrange_relax} can be formulated as a sum of the two following problems

\begin{equation} \label{eq:Lagrangian_subproblem_x}
	\min_{\bm{x}\in\mathbb{B}^{\vert E \vert}}\enskip \sum_{e\in E} (w_Bc_e - \sum_{k\in K} \lambda_e^k)x_e,
\end{equation}
and
\begin{equation} \label{eq:dual_function}
	h(\bm{\lambda}) := \sum_{k\in K} \min_{\bm{y}^k\in \mathcal{P}_k}\enskip (w_Lc_e + \lambda_e^k)y_e^k.
\end{equation}

Assume that we relax $x_e$ to be in $\mathbb{Z}$, then if $(w_Bc_e - \sum_k \lambda_e^k) < 0 \Rightarrow x_e \rightarrow \infty $ and if $(w_Bc_e - \sum_k \lambda_e^k) > 0 \Rightarrow x_e \rightarrow -\infty$ when solving \eqref{eq:Lagrangian_subproblem_x}. Since the Lagrangian dual problem aims to maximize \eqref{eq:Lagrangian_subproblem_x} we can conclude that the optimal dual solution must satisfy
\begin{equation} \label{eq:dual_edge_equality}
	\Omega_e := \{ \bm{\lambda}_e  \mid \sum_{k\in K}\lambda^k_{e} = w_Bc_{e}, \: \bm{\lambda}_e \geq \bm{0}^{\vert K\vert}\}, \enskip e\in E,
\end{equation}
and with these dual constraints, the dual function ($h(\bm{\lambda})$) reduces to \eqref{eq:dual_function}. The dual problem is defined as
\begin{equation} \label{eq:dual_problem}
	\max_{\bm{\lambda}_e\in \Omega_e,\: e\in E} \: h(\bm{\lambda}).
\end{equation}

Note that the Lagrangian subproblem (evaluating $h(\bm{\lambda})$) is a sum of shortest path problems and can be solved efficiently using, for example, Dijkstra's algorithm or A\textsuperscript{*} search algorithm. Also, note that sparsity can be utilized for the subgradients $\bm{\xi}\in \frac{\partial h}{\partial \bm{\lambda}}$, where $\frac{\partial h}{\partial \bm{\lambda}}:=\arg h(\bm{\lambda})$. Since the Lagrangian subproblem has the integrality property we know that the optimal value of the dual problem and the LP relaxation of the CHRP are equal~\cite{moshe1990Relax}.

\section{Cable harness routing algorithm} \label{sec:algorithm}
A graph search algorithm, referred to as the \emph{Harness Routing Heuristic (HRH)}, is explained in Section \ref{sec:harness_heuristic} and searches locally from initial routes. Two methods are used to find initial routes to the HRH: (1) the Lagrangian dual problem is maximized using a projected subgradient method and the Lagrangian subproblem solutions are given to the HRH, see Section \ref{sec:subgrad_method}, and (2) initial routes are heuristically constructed, see Section \ref{sec:aSPHRH}.

We obtain topologically different solutions by locally optimizing various initial routes, as depicted in Figure~\ref{fig:subgradient_method_subproblem_sols}--\ref{fig:subgradient_method_primal_sols}. Our goal is to generate a range of promising topology candidates, as these solutions will undergo further modifications to address more design aspects.

\subsection{The cable harness routing heuristic} \label{sec:harness_heuristic}
\begin{figure}[!ht]
	\centering
	\begin{subfigure}[b]{0.22\textwidth}
		\centering
		\tcbox[size=fbox]{\includegraphics[width=\textwidth]{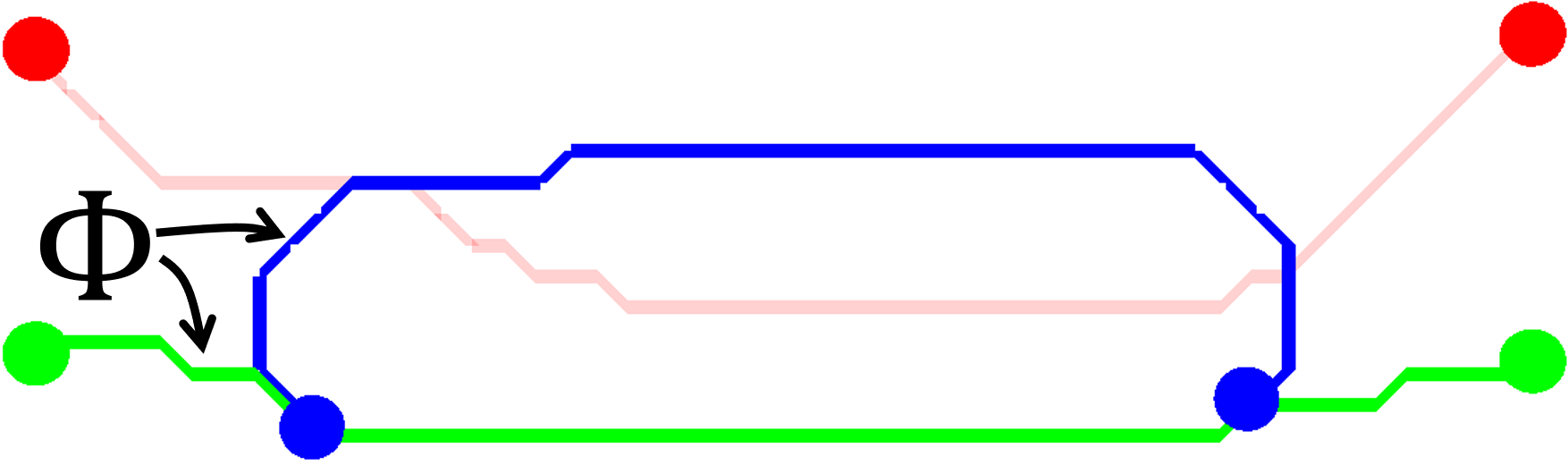}}
		\caption{}
		\label{subfig:local_search_heuristic_a}
	\end{subfigure}
	\hspace{1mm}
	\begin{subfigure}[b]{0.22\textwidth}
		\centering
		\tcbox[size=fbox]{\includegraphics[width=\textwidth]{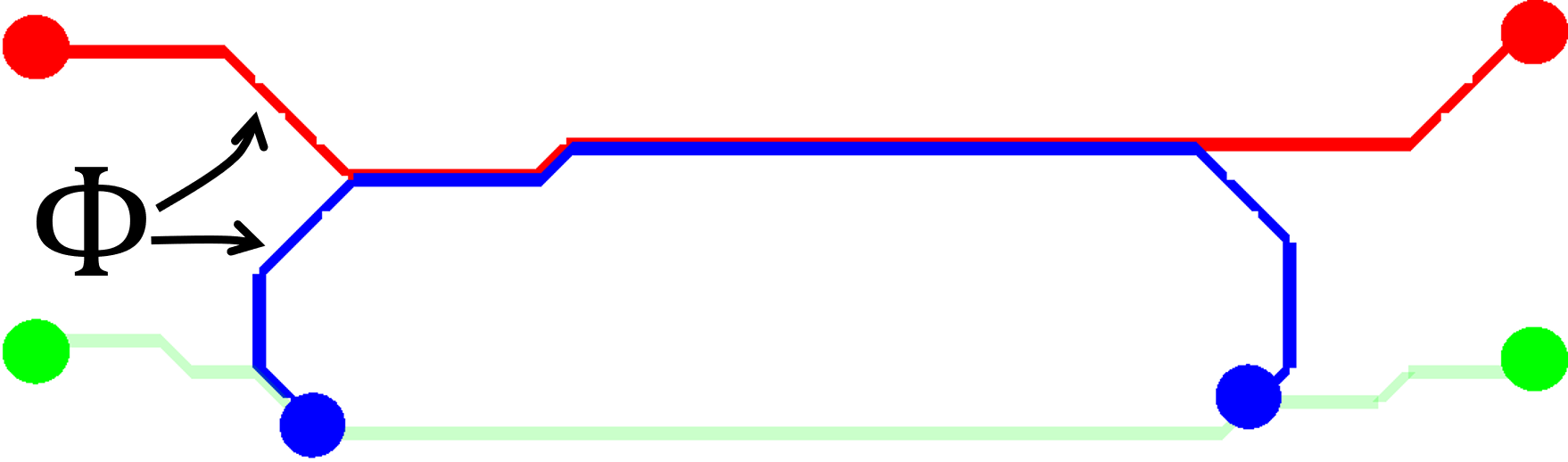}}
		\caption{}
		\label{subfig:local_search_heuristic_b}
	\end{subfigure}
	\begin{subfigure}[b]{0.22\textwidth}
		\centering
		\tcbox[size=fbox]{\includegraphics[width=\textwidth]{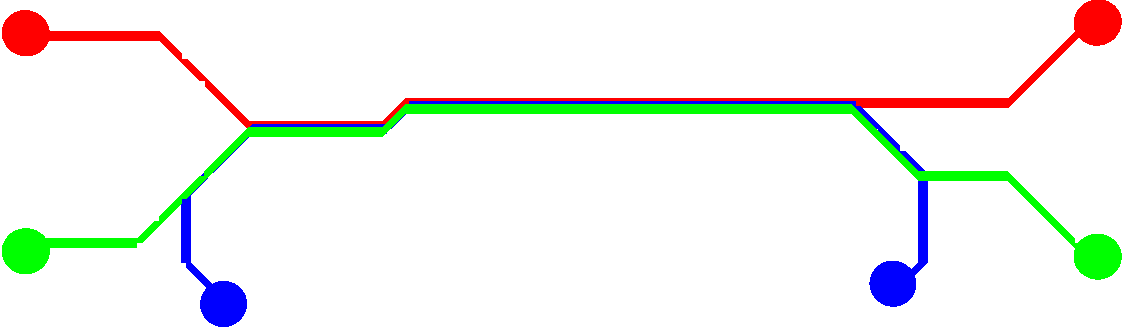}}
		\caption{}
		\label{subfig:local_search_heuristic_c}
	\end{subfigure}
	\hspace{1mm}
	\begin{subfigure}[b]{0.22\textwidth}
		\centering
		\tcbox[size=fbox]{\includegraphics[width=\textwidth]{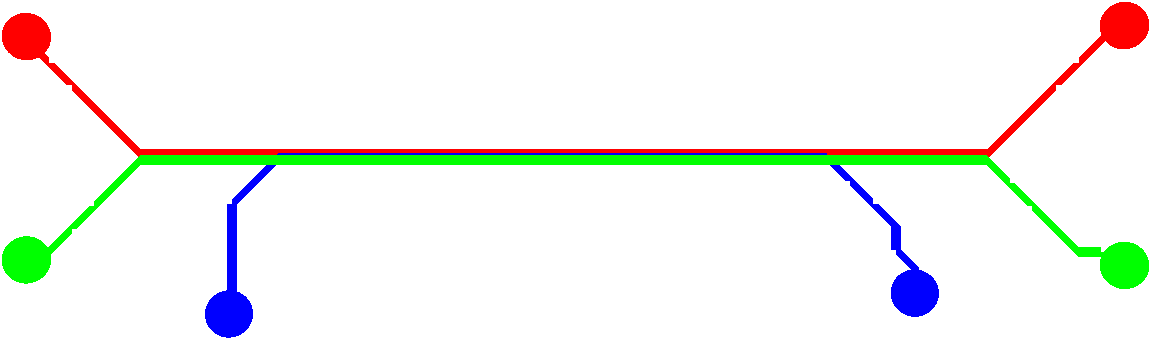}}
		\caption{}
		\label{subfig:local_search_heuristic_d}
	\end{subfigure}
	\caption{Execution of the HRH. (\subref{subfig:local_search_heuristic_a})$\rightarrow$(\subref{subfig:local_search_heuristic_b}): Algorithm \ref{alg:shortest_path_with_edge_activation} applied for the red cable. (\subref{subfig:local_search_heuristic_b})$\rightarrow$(\subref{subfig:local_search_heuristic_c}): Algorithm \ref{alg:shortest_path_with_edge_activation} applied for the green cable. (\subref{subfig:local_search_heuristic_d}): Optimized branch point locations.}
	\label{fig:harness_heuristic}
\end{figure}

The HRH, which is based on Dijkstra's algorithm and  A\textsuperscript{*} search, is summarized by Algorithm \ref{alg:harness_heuristic} and depicted in Figure~\ref{fig:harness_heuristic}. A\textsuperscript{*} search is utilized when we can define a heuristic function that estimates the cost of the cheapest path from any node to a goal node.
\begin{algorithm}
	\textbf{Input:} Initial harness routing.\\
	\textbf{Output:} Optimized harness routing.
	\begin{enumerate}
		\item Apply Algorithm \ref{alg:heuristic_cable_local_search} to iteratively optimize the cables' routes. \label{algstep:harness_heuristic_cable_local_search}
		\item Optimize branch point locations by iteratively optimizing the bundles' and their branches' routes. \label{algstep:harness_heuristic_optimize_branch_locations}
	\end{enumerate}
	\caption{Harness Routing Heuristic (HRH)}
	\label{alg:harness_heuristic}
\end{algorithm} 
Step \ref{algstep:harness_heuristic_cable_local_search} in the HRH is explained with Algorithm \ref{alg:heuristic_cable_local_search} where one cable at a time is routed with respect to other cables' routes (see Figure~\ref{subfig:local_search_heuristic_a}--\ref{subfig:local_search_heuristic_c}) as explained in Algorithm \ref{alg:shortest_path_with_edge_activation}. Algorithm \ref{alg:shortest_path_with_edge_activation} can be seen as solving the CHRP for one cable $\tilde{k}\in K$ where $f_B$ only penalizes traversing of edges not used by the fixed cables in $K\setminus \tilde{k}$. A\textsuperscript{*} search is used in Step \ref{algstep:shortest_path_with_edge_activation_find_sp} in Algorithm \ref{alg:shortest_path_with_edge_activation} by using the shortest-path tree rooted at the end node for $\tilde{k}$ (precomputed through Dijkstra's algorithm with edge costs equal to $w_Lc_e$, $e\in E$) as the heuristic function. In Step \ref{algstep:harness_heuristic_optimize_branch_locations} of the HRH, a bundle and its branching paths are routed to optimality with respect to the CHRP and some fixed topology. This is achieved by summing the shortest distances available from Dijkstra's algorithm for each route connected to the respective branch points, and by a single all-to-all shortest path to find the new branch point locations and the path between them. In all queries of step 2, the  goal node is not known, and hence A\textsuperscript{*} cannot be used.

\begin{algorithm}
	\caption{Cables' route search in a harness}
	\label{alg:heuristic_cable_local_search}
	\begin{algorithmic}
		\State\textbf{Input:} $\bm{y}$ \Comment{Initial routes}
		\State\textbf{Initialize:} $i:=1,\enskip \tilde{k}:=1,\enskip \bm{y}_\text{best}:=\bm{y}$
		\State\textbf{Define:} $f_y := f((\Vert \bm{y}_e \Vert_{\infty})_{e\in E}, \bm{y})$
		\While{$i \leq \vert K\vert$}
		\State $\Phi:=\bigcup_{k\in K\setminus\tilde{k}}\{\bm{y}_\text{best}^k\}$ \Comment{Set of fixed paths}
		\State $\bm{y}^k_\text{new}:=\bm{y}_\text{best}^k, \enskip k\in K\setminus \tilde{k}$
		\State $\bm{y}^{\tilde{k}}_\text{new}$ := Algorithm \ref{alg:shortest_path_with_edge_activation} ($\tilde{k}$, $\Phi$) \Comment{Find new path for $\tilde{k}$}
		\If{$f_y(\bm{y}_\text{new}) < f_y(\bm{y}_\text{best})$}
		\State $\bm{y}_\text{best}:=\bm{y}_\text{new}$ \Comment{Update best solution}
		\State $i:=0$
		\EndIf
		\State $i:=i+1$
		\State $\tilde{k}:=\tilde{k}+1$
		\If{$\tilde{k} > \vert K\vert$}
		\State $\tilde{k}:=1$
		\EndIf
		\EndWhile
		\State return $\bm{y}_\text{best}$
	\end{algorithmic}
\end{algorithm}

\begin{algorithm}
	\textbf{Input:} Cable index $\tilde{k}\in K$. Set of fixed paths $\Phi$. \\
	\textbf{Output:} Optimal path for cable $\tilde{k}$ with respect to \eqref{eq:CHRP} given fixed paths in $\Phi$.
	\begin{enumerate}
		\item Let an edge cost be equal to $w_Lc_e$ if $e\in E$ is traversed by any path in $\Phi$ and equal to $(w_L+w_B)c_e$ otherwise. \label{algstep:shortest_path_with_edge_activation_fix_x_e}
		\item Find the shortest path for $\tilde{k}$ with respect to the edge costs defined in Step \ref{algstep:shortest_path_with_edge_activation_fix_x_e}. \label{algstep:shortest_path_with_edge_activation_find_sp}
	\end{enumerate}
	\caption{Shortest path with non-shared edge penalty}
	\label{alg:shortest_path_with_edge_activation}
\end{algorithm}

\subsection{Subgradient method with the harness routing heuristic} \label{sec:subgrad_method}
To optimize the Lagrangian dual problem \eqref{eq:dual_problem} a projected subgradient method is used in Algorithm \ref{alg:harness_subgrad}, the algorithm is referred to as the \emph{Subgradient Harness Routing Heuristic (SHRH)}. The projection is done edge-wise onto $\Omega_e,\: e\in E$, with the algorithm described in~\cite[p.~461--462]{book_projection}. The step length at iteration $i$ in Step \ref{algstep:harness_subgrad_update_step_and_stop_criteria} is updated according to the rule in~\cite{fisher1981lagrangian}:

\begin{equation*}
	\eta^{[i]}:=\delta^{[i]} \frac{\hat{h}-h(\bm{\lambda}^{[i]})}{\Vert \bm{\xi} \Vert^2}, \enskip 0 < \delta^{[i]} \leq 2,
\end{equation*}

where $\hat{h}$ is the best known upper bound on $h$.

\begin{figure*}[!ht]
	\centering
	\begin{subfigure}[t]{0.30\textwidth}
		\centering
		\includegraphics[width=\textwidth,trim={1.4cm 0 1.7cm 0},clip ]{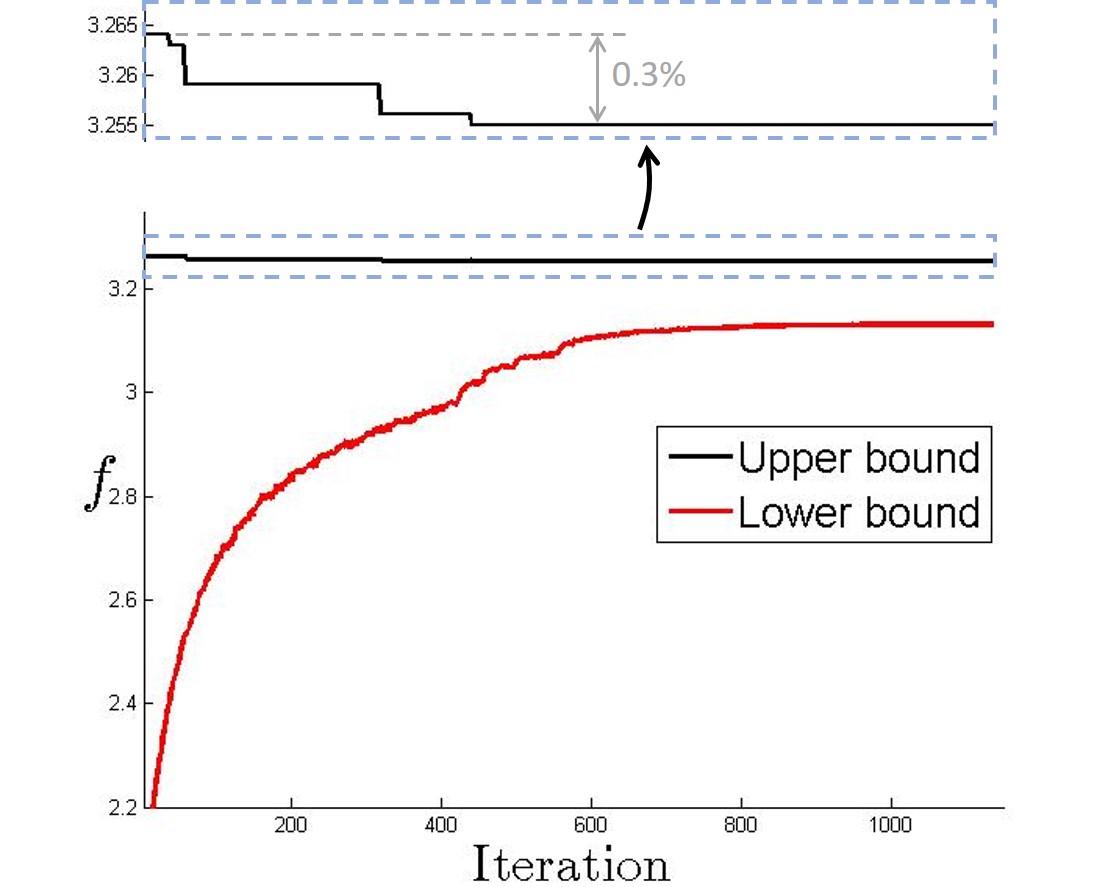}
		\caption{Run the subgradient algorithm to maximize the Lagrangian dual problem and get a lower bound. Data example from solving Case B in Section \ref{sec:industrial_cases} with $i_{\text{HRH}}=10$.}
		\label{fig:subgradient_method_lb_ub}
	\end{subfigure}
	\quad
	\begin{subfigure}[t]{0.30\textwidth}
		\centering
		\begin{tabular}[b]{c}
			\includegraphics[width=\textwidth,height=3cm]{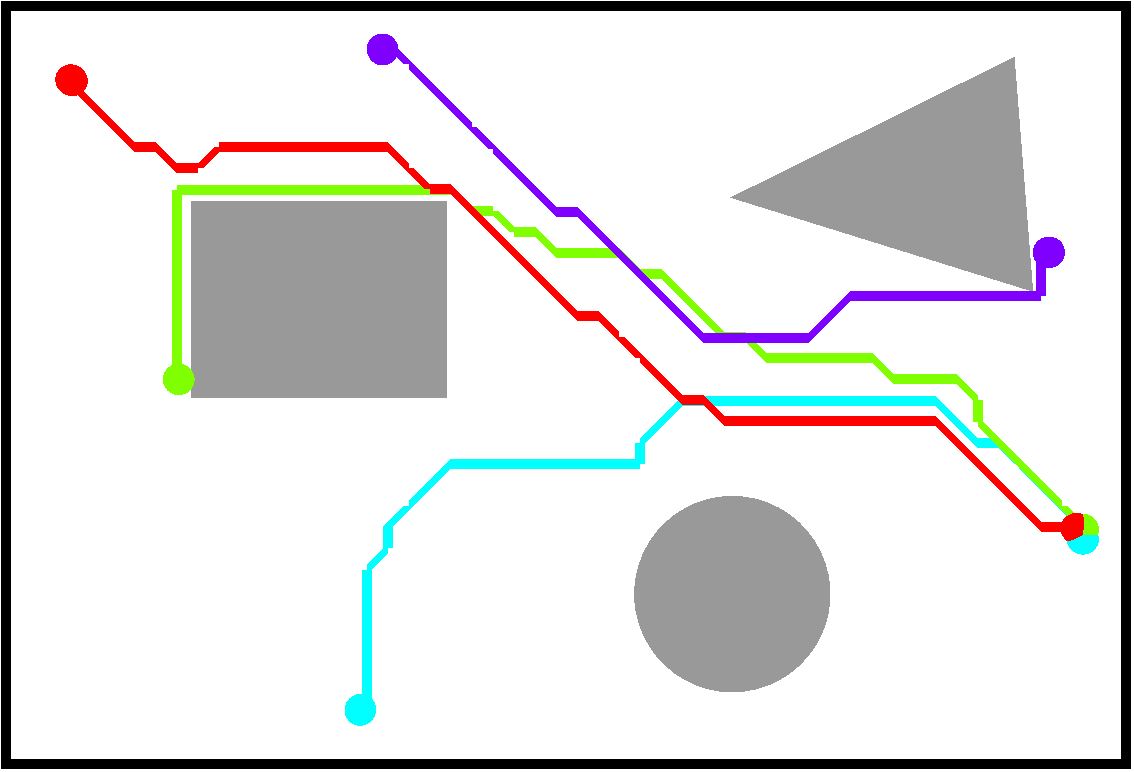} \\
			\includegraphics[width=\textwidth,height=3cm]{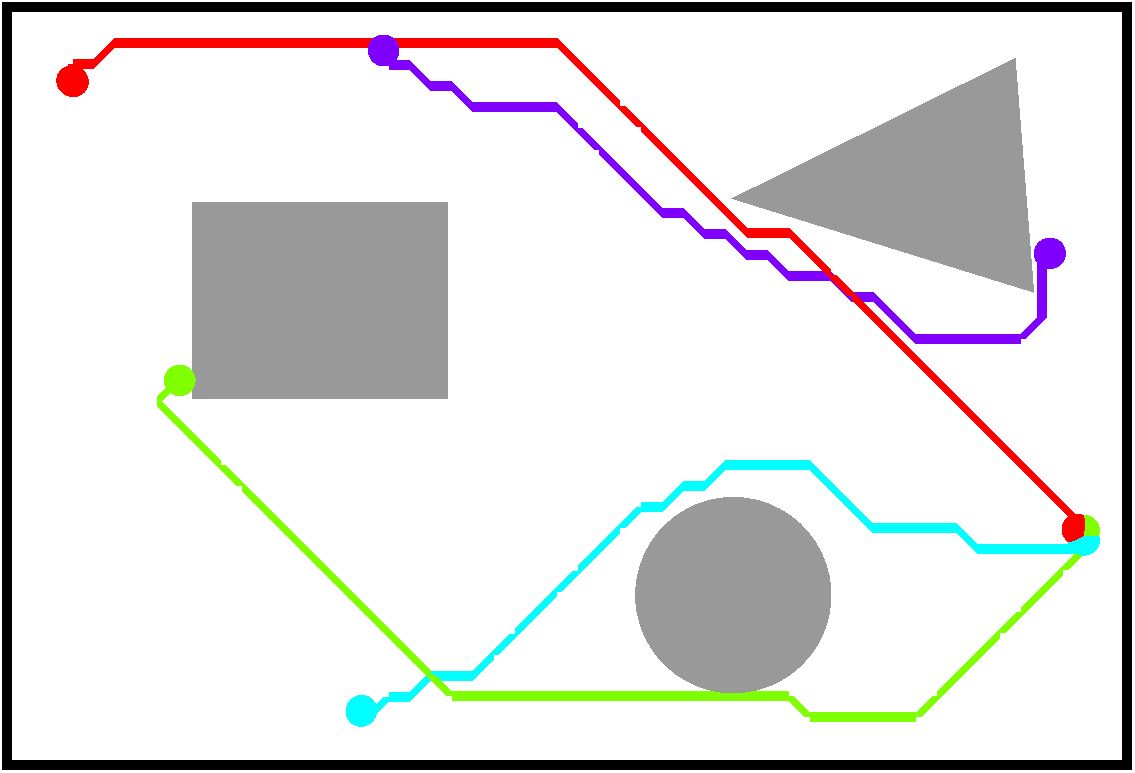}
		\end{tabular}
		\caption{Lagrangian subproblem solutions are obtained throughout the subgradient iterations.}
		\label{fig:subgradient_method_subproblem_sols}
	\end{subfigure}
	\quad
	\begin{subfigure}[t]{0.30\textwidth}
		\centering
		\begin{tabular}[b]{c}
			\includegraphics[width=\textwidth,height=3cm]{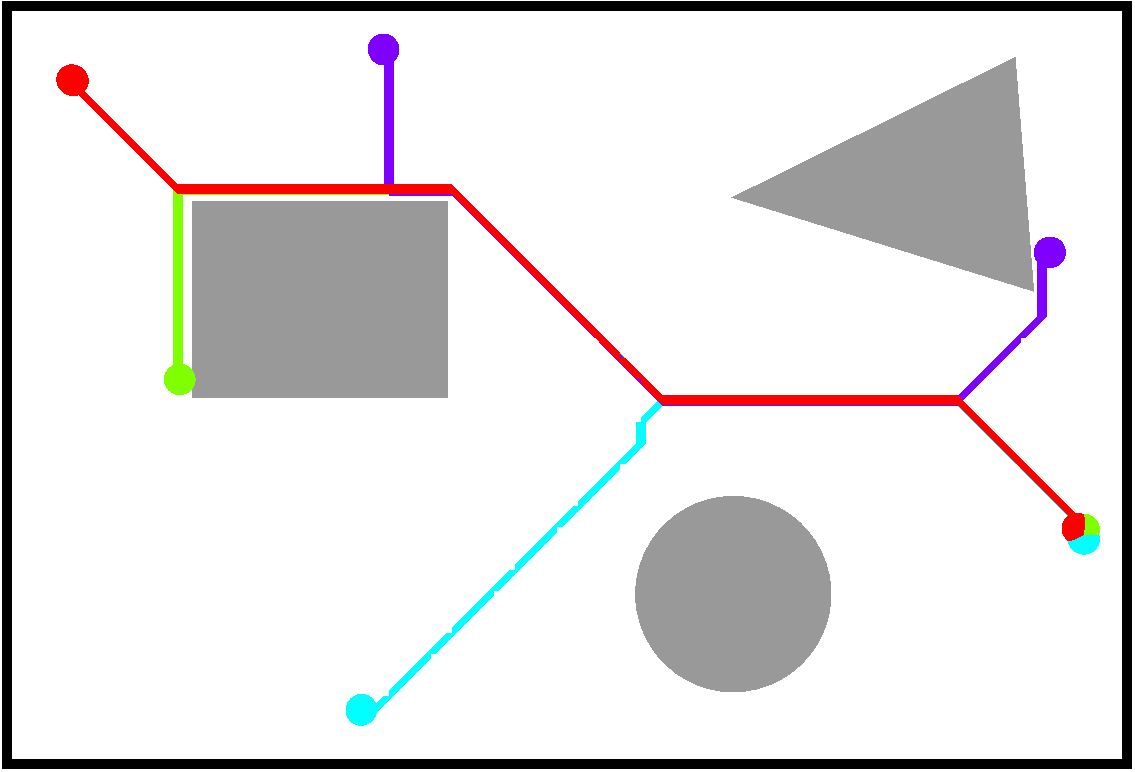} \\
			\includegraphics[width=\textwidth,height=3cm]{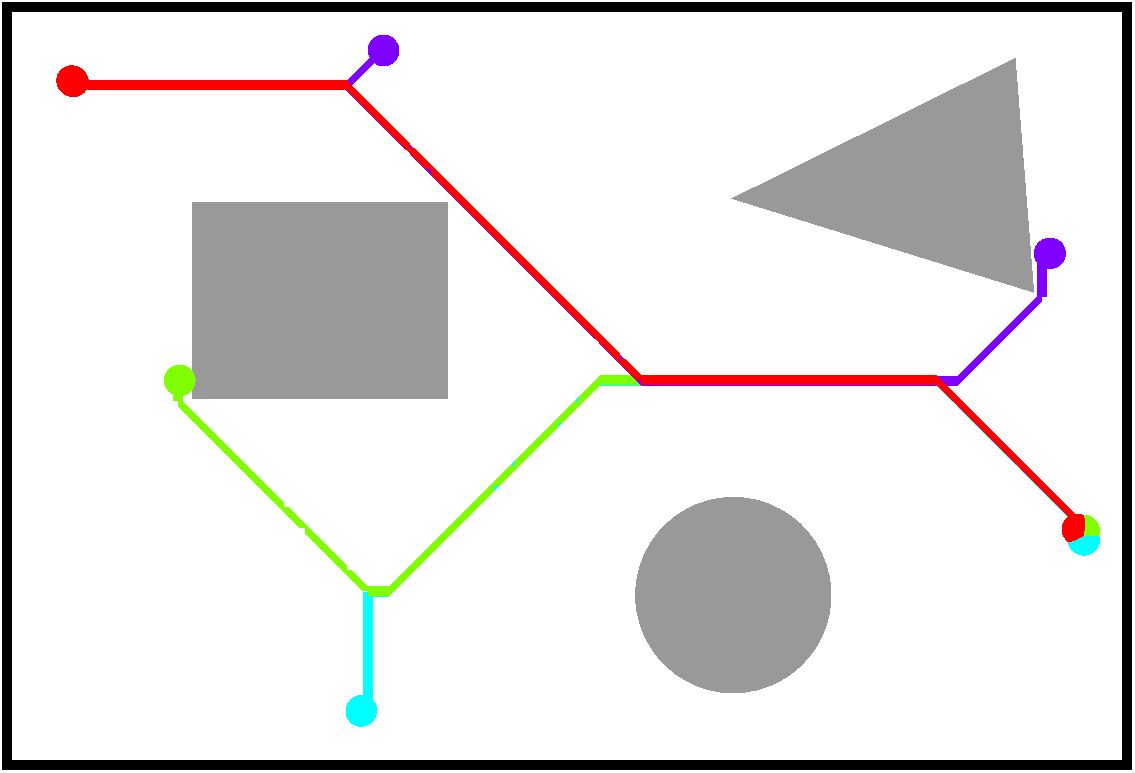}
		\end{tabular}
		\caption{Given initial subproblem solutions, the HRH is executed which yields an upper bound and topologically different local optimal solutions.}
		\label{fig:subgradient_method_primal_sols}
	\end{subfigure}
	\caption{Depiction of the SHRH.}
	\label{fig:subgradient_method}
\end{figure*}

The process of maximizing the dual problem and using the Lagrangian subproblem solutions as input to the HRH is depicted in Figure~\ref{fig:subgradient_method} where we can see how different initial routings result in topologically different local optimal primal solutions.

\begin{algorithm}
	\begin{enumerate}
		\item Define $i_{\text{HRH}},i_{\text{stag}}\in \mathbb{N}$, $\epsilon\in\mathbb{R}_+$, and $\delta^{[0]}\in (0,2]$. Set $i:=0$ and choose an initial point $\bm{\lambda}^{[0]}\in\prod_{e\in E} \Omega_e$.
		\item Compute a subgradient $\bm{\xi}\in \frac{\partial h}{\partial \bm{\lambda}^{[i]}}$ and update the dual variables as
		\begin{equation*}
			\bm{\lambda}_e^{[i+1]} := P_{\Omega_e}\big( \bm{\lambda}_e^{[i]} + \eta^{[i]}\bm{\xi}_e \big), \enskip e\in E,
		\end{equation*}
		with step length $\eta^{[i]}\in \mathbb{R}_+$, where $P_{\Omega_e}(\cdot)$ is a projection operator on the set $\Omega_e$. \label{algstep:harness_subgrad_update_dual}
		\item Execute the HRH (Algorithm \ref{alg:harness_heuristic}) given a subproblem solution if $i$ is divisible by $i_{\text{HRH}}$.
		\item Set $i:=i+1$, update $\eta^{[i]}$ and go to Step \ref{algstep:harness_subgrad_update_dual} until some stopping criterion is met. \label{algstep:harness_subgrad_update_step_and_stop_criteria}
	\end{enumerate}
	\caption{Subgradient Harness Routing Heuristic (SHRH)}
	\label{alg:harness_subgrad}
\end{algorithm}

We use the following criterion to determine when to terminate the SHRH due to stagnation (when $i\geq i_{\text{stag}}$)

\begin{equation*}
	\frac{h^* - h(\bm{\lambda}^{[i-i_{\text{stag}}]})}{h(\bm{\lambda}^{[i-i_{\text{stag}}]})} < \epsilon, \enskip \epsilon > 0, \enskip i_{\text{stag}}\in \mathbb{N},
\end{equation*}

where $h^*$ is the best known value of $h$. This is computationally unnecessary if the goal is solely to obtain a low upper bound, without the need for a high lower bound. From Figure~\ref{fig:subgradient_method_lb_ub} we can see that low upper bounds can be obtained in early iterations, motivating us to terminate the SHRH earlier (e.g., when a desired number of candidate solutions has been obtained). 

\subsection{$\alpha$-shortest paths with the harness routing heuristic} \label{sec:aSPHRH}
The approach presented in this section, referred to as the \emph{$\alpha$-Shortest Paths Harness Routing Heuristic ($\alpha$-SPHRH)}, provides an alternative method, relative to the SHRH, for obtaining topologically different initial routings. The main difference between the SHRH and the $\alpha$-SPHRH is how the input to the HRH is found.

The set of \emph{$\alpha$-shortest paths} for a cable consists of unique paths that are no longer than a factor $\alpha > 1$ multiplied by the length of the shortest path. The method of finding $\alpha$-shortest paths in an early stage where more design factors are to be considered (and not just the length of the path) has been addressed before in~\cite{hermansson2021routing}. 

The first step in the $\alpha$-SPHRH is to apply Algorithm 2 in~\cite{hermansson2021routing} to generate sets of $\alpha$-shortest paths $\Phi^{\alpha}_k$, $k\in K$. We let $\vert\Phi^{\alpha}_k\vert=n_{\Phi}$, $k\in K$, where $n_{\Phi}\in\mathbb{N}$ is a parameter and the elements in $\Phi^{\alpha}_k$ are spatially distinct. The following steps explain how to get an initial solution:
\begin{enumerate}
	\item Let $\Phi_k:=\Phi^{\alpha}_k, \enskip k\in K$. Determine a unique sequence of all cables.
	\item For each consecutive element $\tilde{k}$ in the sequence, set $\Phi_{\tilde{k}}:=\emptyset$ and apply Algorithm \ref{alg:shortest_path_with_edge_activation} with $\tilde{k}$ and $\cup_{k\in K}\Phi_k$ as input. Add the resulting path to $\Phi_{\tilde{k}}$ ($\Rightarrow \vert \Phi_{\tilde{k}} \vert=1$).
	\item Save the initial solution $\cup_{k\in K}\Phi_k$.
\end{enumerate}
Repeat the steps above for a new unique sequence of cables to get another initial solution until the number of solutions is enough, and then run the HRH for each one.

The primary advantage of the $\alpha$-SPHRH over the SHRH is its reduced computation time for finding initial routings. However, a drawback is that using these rule-based constructed initial routes can lead to omission of certain local optima in specific cases, which the SHRH is capable of finding.

\section{Results} \label{sec:results}
The algorithms have been implemented in C\texttt{++} in a prototype version of the IPS software~\cite{ips}. All tests were run on a computer with an Intel® Core™ i7-6700K processor (4.00GHz) and 32.0 GB RAM. 

The parameters for the SHRH were set as the following: $i_{\text{HRH}}:=25$, $i_{\text{stag}}:=100$, $\epsilon:=10^{-4}$ and $\delta^{[0]}:=1.5$. $\delta^{[i]}$ was multiplied with 0.8 every time the dual objective function value did not improve within 10 iterations. Given the selected values of $i_{\text{stag}}$ and $\epsilon$, we are confident that stagnation will be reached. It is significantly more computationally efficient to take a subgradient step than executing the HRH, which motivates us to avoid using a low value for $i_{\text{HRH}}$. The remaining parameters were determined by running the algorithm with various settings. This was done on instances not covered in this section. The selected setting resulted in a fast convergence rate and a high lower bound. The initial feasible dual solution was set as $\lambda_e^k:=w_Bc_e/\vert K\vert,\:k\in K,\:e\in E$. The parameters for the $\alpha$-SPHRH were set as the following: $\alpha:=1.2$, $n_{\Phi}:=7$, and the number of constructed initial solutions was equal to 5. These settings enabled the $\alpha$-SPHRH to find a variety of local optimal solutions and an approximated Pareto front close to the one given by the SHRH. Both the SHRH and $\alpha$-SPHRH were parallelized over a set of bundle weights, uniformly distributed within some interval.

Most instances used in this section, such as those shown in Figure~\ref{fig:many_cables_scene} and \ref{fig:processing_pre}, do not represent real-world cases. However, they were generated to resemble typical harness routing problems by organizing the start nodes into clusters and the end nodes into separate clusters, while maintaining a certain distance between them. Figure~\ref{fig:case_A-C_manual_routing} depicts three instances based on real-world cases from the automotive industry, where manually created harness solutions were provided, and the terminals were placed at connector positions.

In the following subsections, we investigated the following:
\begin{enumerate}
	\item Comparison of the SHRH against an exact solver.
	\item Computational complexity for increasing number of decision variables.
	\item Comparison of the SHRH and $\alpha$-SPHRH with respect to computation time and objective function value.
	\item Further processing using the grid graph and the use of other methods to fulfil certain design aspects.
	\item Comparison between this paper's approach and methods used in related work.
\end{enumerate}

\subsection{A comparison of optimal values for large duality gaps} \label{sec:exact_solver}
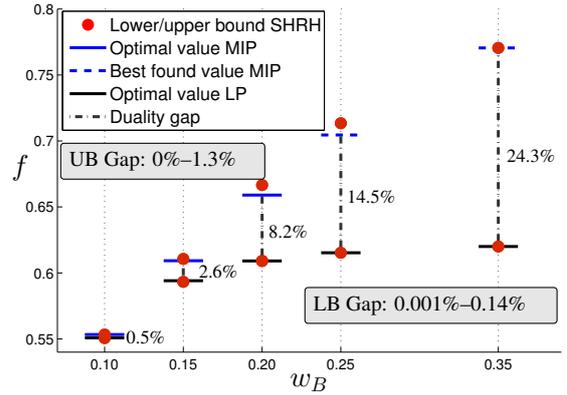
\begin{figure}[!ht]
	\centering
	\begin{subfigure}[b]{0.45\textwidth}
		\begin{overpic}[trim={1cm 0 1cm 0},clip,width=\textwidth]{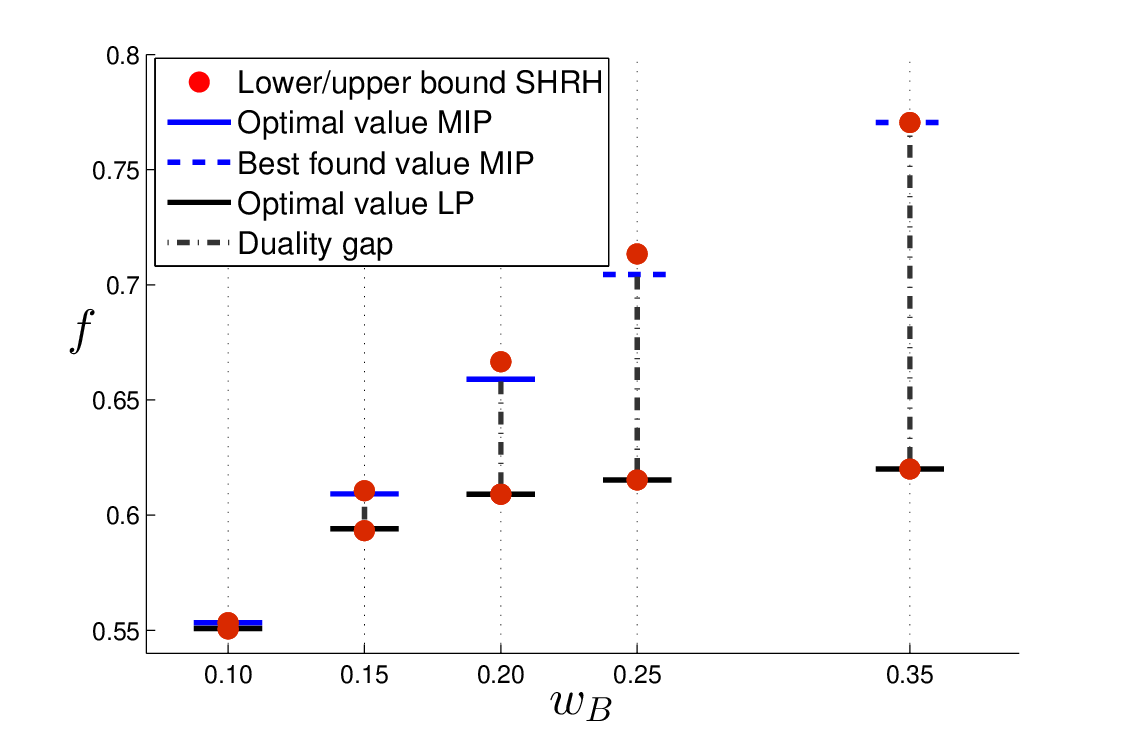}
			\put(8,38){
				\tcbox[size=fbox,colback=lightlightgray]{
					\begin{minipage}{7em}
						{\footnotesize UB Gap: 0\%--1.3\%}
					\end{minipage}
				}
			}
			\put(50,13){
				\tcbox[size=fbox,colback=lightlightgray]{
					\begin{minipage}{8.8em}
						{\footnotesize LB Gap: 0.001\%--0.14\%}
					\end{minipage}
				}
			}
			\put(20.5, 9.7){\scriptsize 0.5\%}
			\put(33, 21){\scriptsize 2.6\%}
			\put(45, 28){\scriptsize 8.2\%}
			\put(58.5, 34){\scriptsize 14.5\%}
			\put(86, 41){\scriptsize 24.3\%}		
		\end{overpic}
	\end{subfigure}
	\caption{Lower and upper bound from an LP and MIP solver and the SHRH. The HRH can find near-optimal solutions.}
	\label{fig:big_duality_gaps}
\end{figure}

In this section, a \emph{Mixed-Integer Programming (MIP)} solver and an LP solver from Gurobi~\cite{gurobi} are used to solve the CHRP and the LP-relaxed version of it. The duality gap is computed relatively as $(f^* - h^*) / h^*$, where $f^*$ and $h^*$ are the best known values of $f$ and $h$, respectively. For some problems, there is a low optimal duality gap, and it is possible to prove that near-optimality has been reached with the SHRH, as for Case A in section \ref{sec:industrial_cases}. For cases with a big duality gap (due to weak lower bounds), the gap is not a suitable measurement for optimality. In Figure~\ref{fig:big_duality_gaps} it is confirmed that the SHRH reaches near-optimality for both the primal and dual problem for a small instance with big duality gaps for different bundle weights. All bounds from the MIP solver are optimal except for the upper bound for $w_B=0.25$ and $w_B=0.35$; the solver ran for six respective twelve days to obtain these objective function values with a proven relative optimality gap equal to 2.15\% respective 5.17\%. The result in Figure~\ref{fig:big_duality_gaps} is an indication that although the duality gap is big and cannot be used as an optimality measurement, we still get good solutions. UB Gap in Figure~\ref{fig:big_duality_gaps} shows the range of the relative gap between the upper bounds of the two solvers for all weights, computed as $(\text{UB}_\text{SHRH}-\text{UB}_\text{MIP}) / \text{UB}_\text{MIP}$, and similar for the lower bounds (LB Gap).

It has been observed that the duality gap increases with a positive correlation with respect to $w_B$ as in Figure~\ref{fig:big_duality_gaps} and \ref{fig:case_A-C_lb_ub_vs_weights}. This correlation is expected, as the optimal objective function values of the dual problem and the LP relaxation of the CHRP are equal. The CHRP is an LP problem for $w_B=0$, but this is not the case for $w_B>0$.

\subsection{Computational complexity} \label{sec:computational_complexity}
\begin{figure}[!ht]
	\centering
	\begin{subfigure}[b]{0.45\textwidth}
		\centering
		\includegraphics[width=\textwidth]{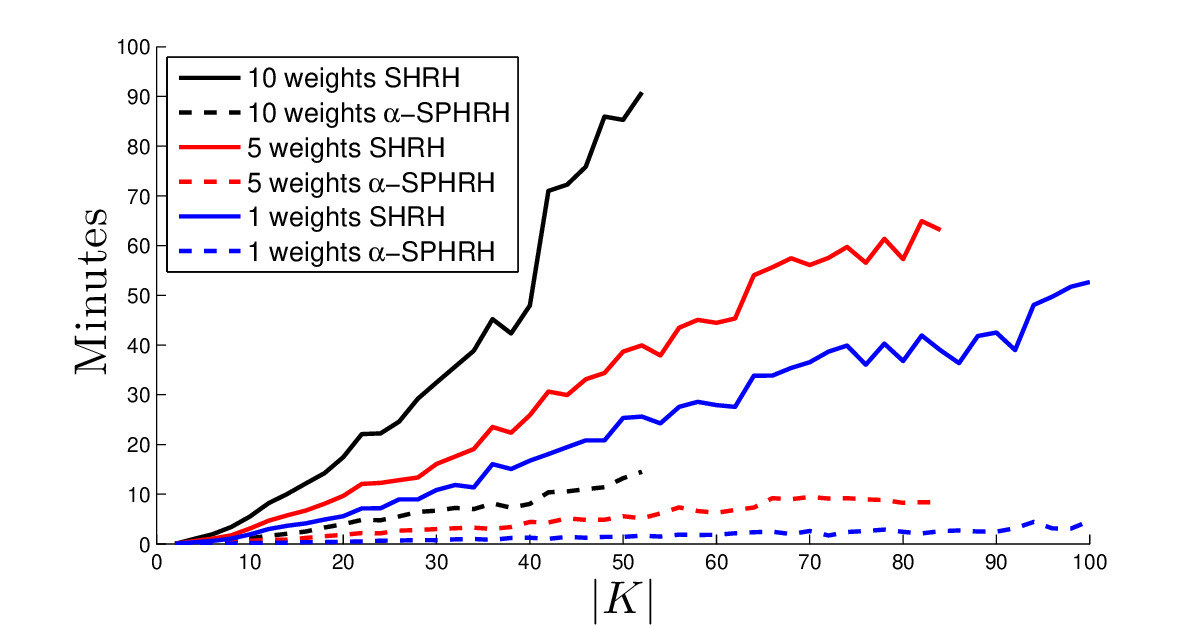}
		\caption{Linear-linear plot.}
		\label{fig:many_cables_time}
	\end{subfigure}
	\begin{subfigure}[b]{0.45\textwidth}
		\centering
		\includegraphics[width=\textwidth]{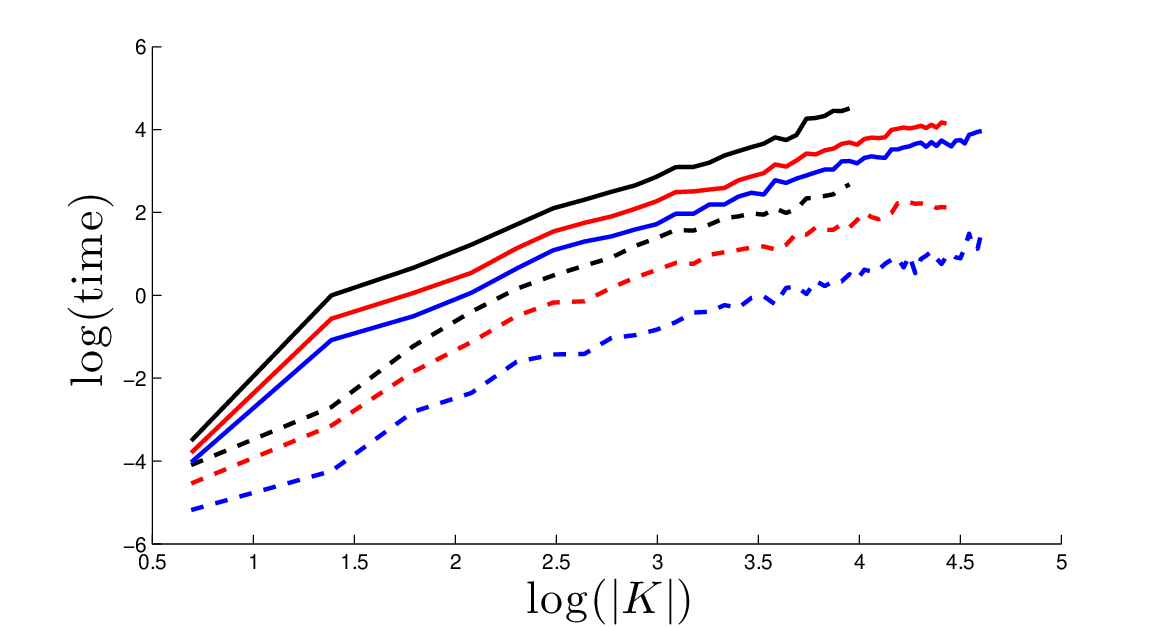}
		\caption{Log-log plot.}
		\label{fig:many_cables_log_time}
	\end{subfigure}
	\begin{subfigure}[b]{0.4\textwidth}
		\centering
		\begin{overpic}[width=\textwidth]{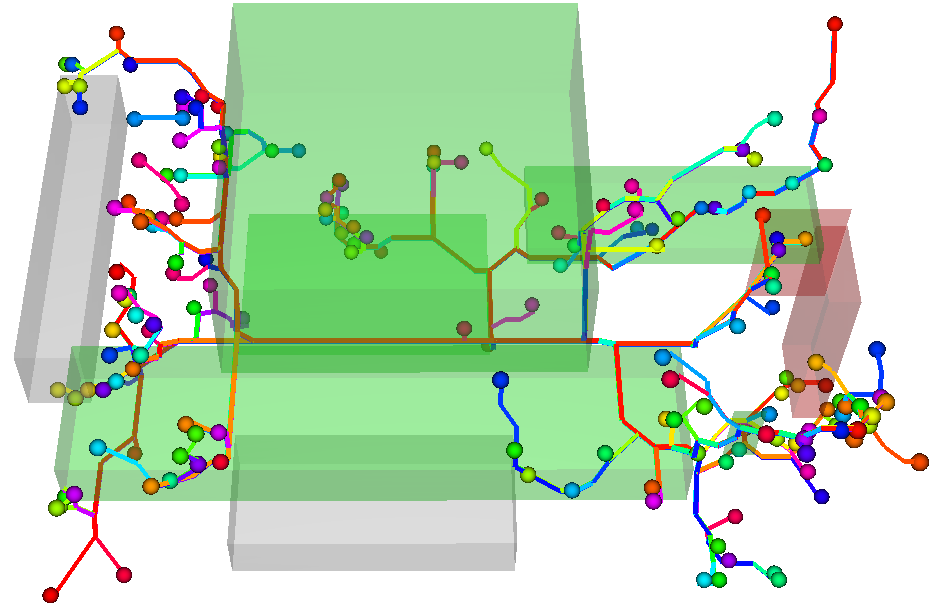}
			\put(33,5){$\vert K\vert=100$}
		\end{overpic}
		\caption{A solution of the case with 100 cables.}
		\label{fig:many_cables_scene}
	\end{subfigure}
	\caption{Computation time for increasing number of cables and a different number of bundle weight values, both for SHRH and $\alpha$-SPHRH. Settings: $\enskip \vert V\vert=20,944$.}
	\label{fig:many_cables}
\end{figure}

\begin{figure}[!ht]
	\centering
	\begin{subfigure}[b]{0.23\textwidth}
		\centering
		\includegraphics[width=\textwidth]{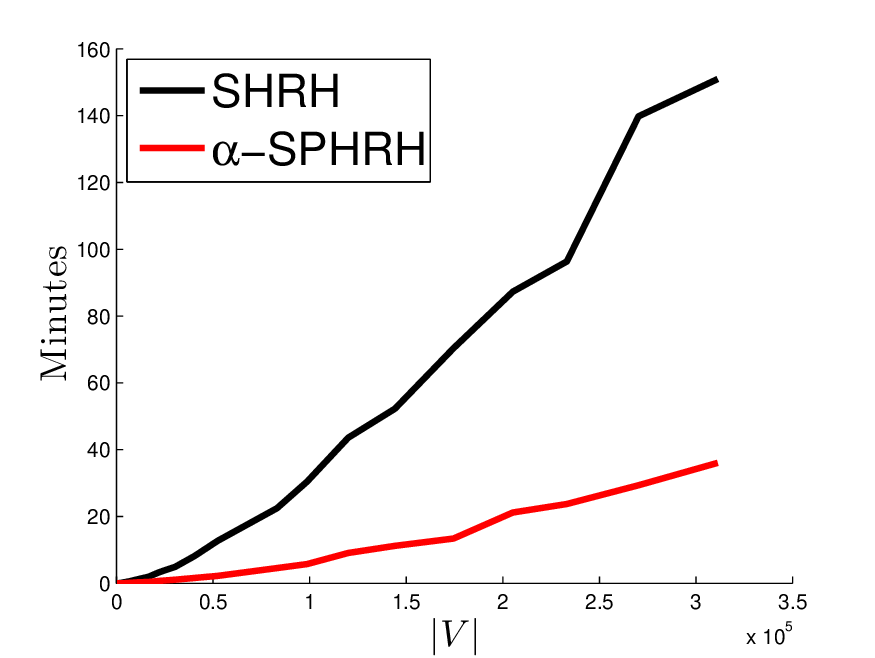}
		\caption{Linear-linear plot.}
		\label{subfig:increasing_grid_time}
	\end{subfigure}
	\begin{subfigure}[b]{0.23\textwidth}
		\centering
		\includegraphics[width=\textwidth]{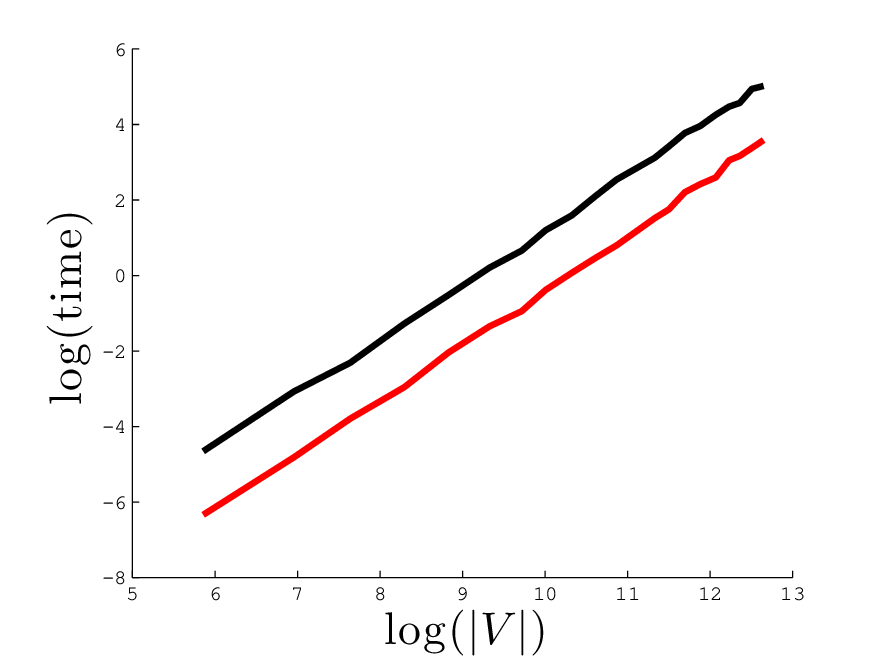}
		\caption{Log-log plot.}
		\label{subfig:increasing_grid_log_time}
	\end{subfigure}
	\caption{Computation time for increasing grid resolution. Settings: $\vert K\vert = 10$ and parallelized over five values of $w_B$.}
	\label{fig:increasing_grid}
\end{figure}

Computation times for increasing number of cables are shown in Figure~\ref{fig:many_cables} and for increasing grid resolution (by reducing the cell size) in Figure~\ref{fig:increasing_grid}. In Figure~\ref{fig:many_cables_time} and \ref{subfig:increasing_grid_time}, we can observe the difference in computation time between SHRH and $\alpha$-SPHRH. The latter is faster due to fewer executions of the HRH and no need to solve Lagrangian subproblems. In Figure~\ref{fig:many_cables_log_time} and \ref{subfig:increasing_grid_log_time}, there is a polynomial trend in computational time for increasing number of decision variables. For this case, the time complexities, based on the numeric experiments, are approximately $O({\vert K\vert}^{2})$ and $O({\vert V\vert}^{1.5})$. Figure~\ref{fig:many_cables} also illustrates the considerable computational resources required when dealing with numerous objective function weights. The observed polynomial trend is interesting as it is challenging to provide a formal proof. HRH is a local search method with a neighbourhood that can be explored in polynomial time, hence it falls under the category of polynomial local search problems (PLS)~\cite{papadimitriou1990complexity}. PLS algorithms do not necessarily have to be polynomial, as it is uncertain whether they will require numerous small improvements. In practice, the improvements can be large and few enough, leading to polynomial time performance, as in Figures~\ref{fig:many_cables_log_time} and \ref{subfig:increasing_grid_log_time}. 

\subsection{Industrial cases} \label{sec:industrial_cases}
\begin{figure}[!ht]
	\centering
	\begin{subfigure}[c]{0.15\textwidth}
		\centering
		\begin{overpic}[trim={0 0 3cm 0},clip,width=\textwidth]{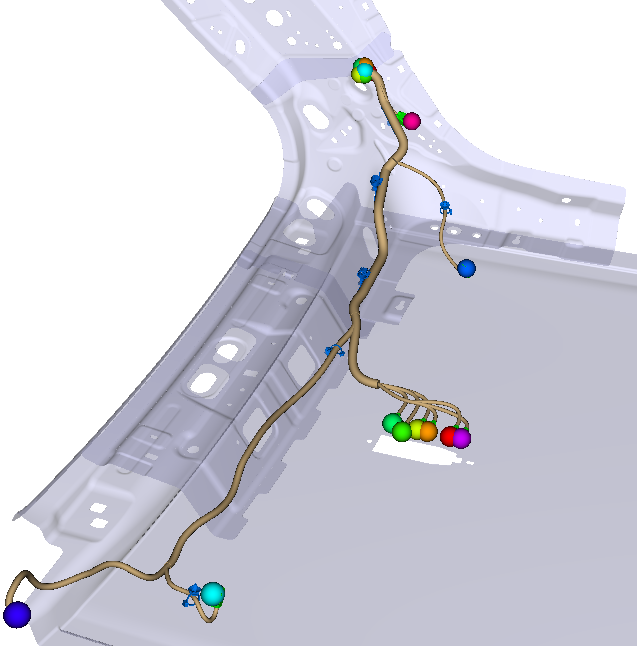}
			\put(40,10){$\vert K\vert=10$}
		\end{overpic}
		\caption{Case A (roof).}
		\label{fig:case_A_manual_routing}
	\end{subfigure}	
	\begin{subfigure}[c]{0.23\textwidth}
		\centering
		\begin{overpic}[trim={0 0 0.5cm 0},clip,width=\textwidth,height=3cm]{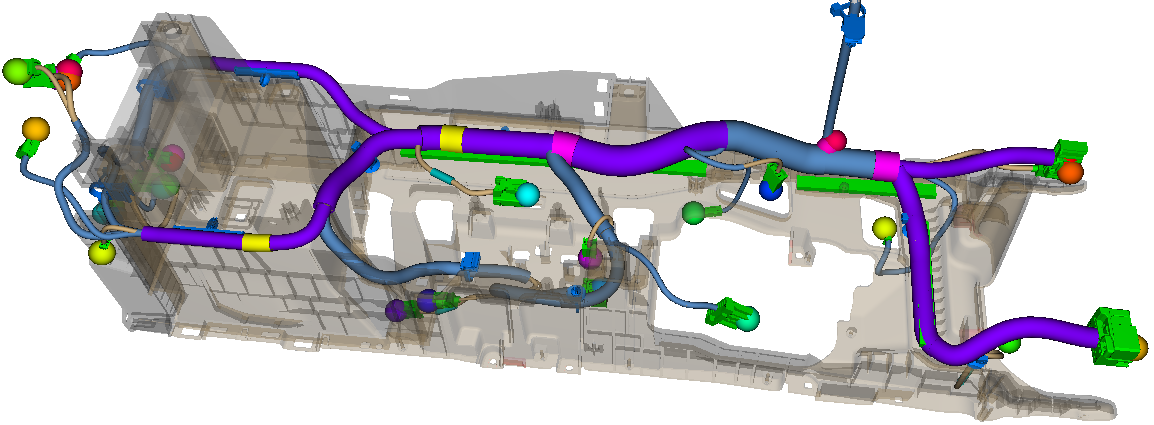}
			\put(25,1){$\vert K\vert=17$}
		\end{overpic}
		\caption{Case B (tunnel console).}
		\label{fig:case_B_manual_routing}
	\end{subfigure}
	\begin{subfigure}[c]{0.23\textwidth}
		\centering
		\begin{overpic}[width=\textwidth]{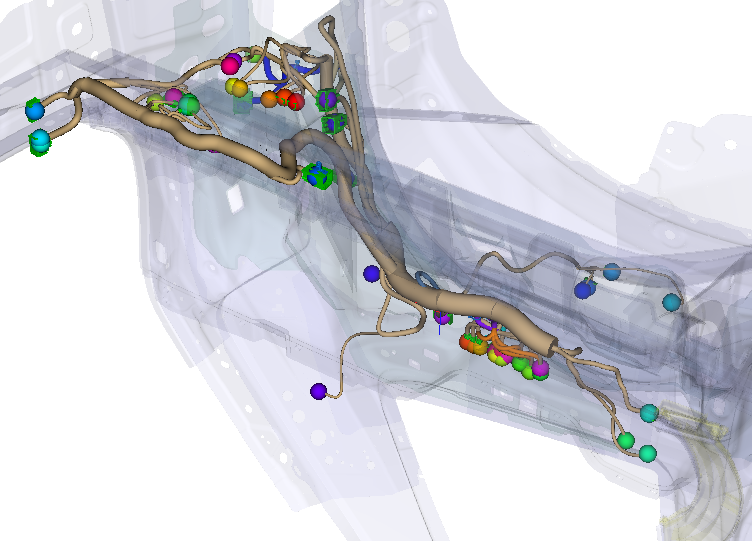}
			\put(30,10){$\vert K\vert=22$}
		\end{overpic}
		\caption{Case C (door).}
		\label{fig:case_C_manual_routing}
	\end{subfigure}
	\caption{Manually made harness designs of the three cases from the automotive industry. Courtesy of Volvo Car Corporation.}
	\label{fig:case_A-C_manual_routing}
\end{figure}

The main purposes of the industrial cases are to test the computation time for industrial-sized instances and compare the best objective function values between the SHRH and $\alpha$-SPHRH. Each instance is solved and parallelized over ten values of $w_B$. The design of the cost field is kept very simple and the resulting solutions in other design aspects than the objective function value are not analysed. Manually made routings for Case A--C can be seen in Figure~\ref{fig:case_A-C_manual_routing}, and the number of cables and grid resolution for each case can be seen in Table~\ref{tab:industrial_cases_numcables_gridsize}. 

\begin{table}[!ht]
	\centering
	\caption{Number of cables and grid resolution for Case A--C.}
	\begin{tabular}{ lcl } 
		Case & $\vert K\vert$ & \multicolumn{1}{c}{$\vert V \vert$} \\ 
		\hline
		A & 10	 	&   $54\times 17 \times 12 = 11,016$ \\ 
		B & 17 	&  $54\times 15 \times 14 = 11,340$  \\
		C & 22 	&  $65\times 42 \times 37 = 101,010$ 	\\
		\hline
	\end{tabular}
	\label{tab:industrial_cases_numcables_gridsize}
\end{table}

\begin{figure}[!ht]
	\centering
	\begin{subfigure}[b]{0.45\textwidth}
		\centering
		\begin{overpic}[width=\textwidth]{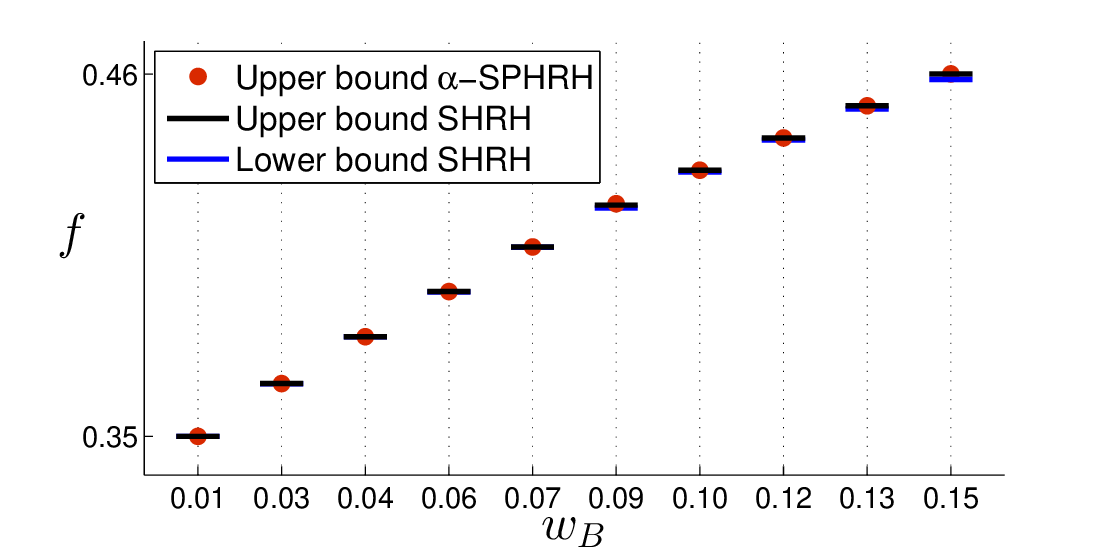}
			\put(36,10){
				\tcbox[size=fbox]{
					\begin{minipage}{10.3em}
						{Duality gap: 0\%--0.36\%}
					\end{minipage}
				}
			}
		\end{overpic}
		\caption{Case A.}
		\label{fig:case_A_weight_vs_ub/lb}
	\end{subfigure}
	\begin{subfigure}[b]{0.45\textwidth}
		\centering
		\begin{overpic}[width=\textwidth]{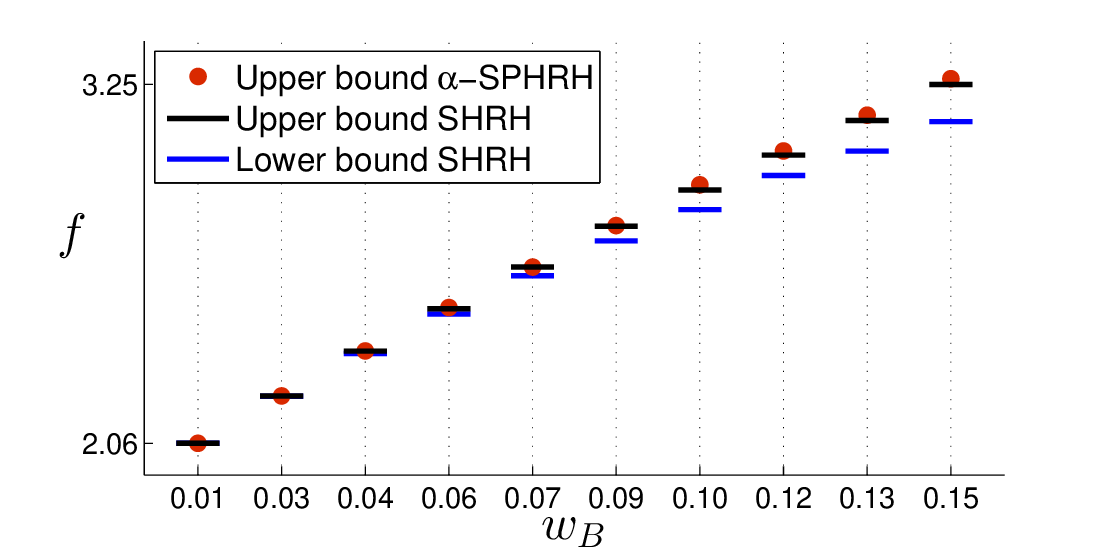}
			\put(36,10){
				\tcbox[size=fbox]{
					\begin{minipage}{11.5em}
						{Duality gap: 0.01\%--3.97\%}
					\end{minipage}
				}
			}
		\end{overpic}
		\caption{Case B.}
		\label{fig:case_B_lb_ub_vs_weights}
	\end{subfigure}
	\begin{subfigure}[b]{0.45\textwidth}
		\centering
		\begin{overpic}[width=\textwidth]{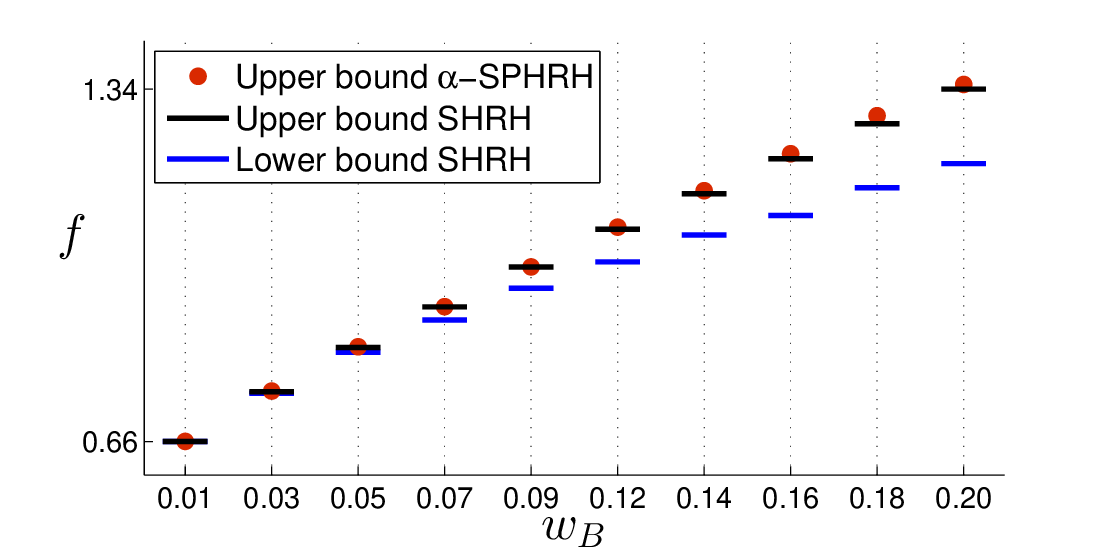}
			\put(36,10){
				\tcbox[size=fbox]{
					\begin{minipage}{12.1em}
						{Duality gap: 0.01\%--11.90\%}
					\end{minipage}
				}
			}
		\end{overpic}
		\caption{Case C.}
		\label{fig:case_C_lb_ub_vs_weights}
	\end{subfigure}
	\caption{Best upper and lower bound found from the SHRH and best upper bound from the $\alpha$-SPHRH for increasing $w_B$. The ranges of the increasing relative duality gaps are presented in percentages.}
	\label{fig:case_A-C_lb_ub_vs_weights}
\end{figure}

The $\alpha$-SPHRH was significantly faster than the SHRH, gave fewer candidate solutions (see Table~\ref{tab:industrial_cases_time_numsols}), and resulted in equally, or slightly worse, best found objective function values (see Figure~\ref{fig:case_A-C_lb_ub_vs_weights}). It is not only the best candidate solution for each bundle weight that is interesting since it might not be the best solution after further optimization over other design factors, therefore the number of (promising and topologically different) candidates is important.

The computation time for the $\alpha$-SPHRH is considered to be acceptable for these large-scale 3D industrial instances. The SHRH can be terminated much earlier (and therefore vastly reduce the computation time) since promising primal solutions are often found early as in Figure~\ref{fig:subgradient_method_lb_ub}---the question is how many candidate solutions we want.

\begin{figure}[!ht]
	\centering
	\begin{subfigure}[b]{0.45\textwidth}
		\centering
		\includegraphics[width=\textwidth]{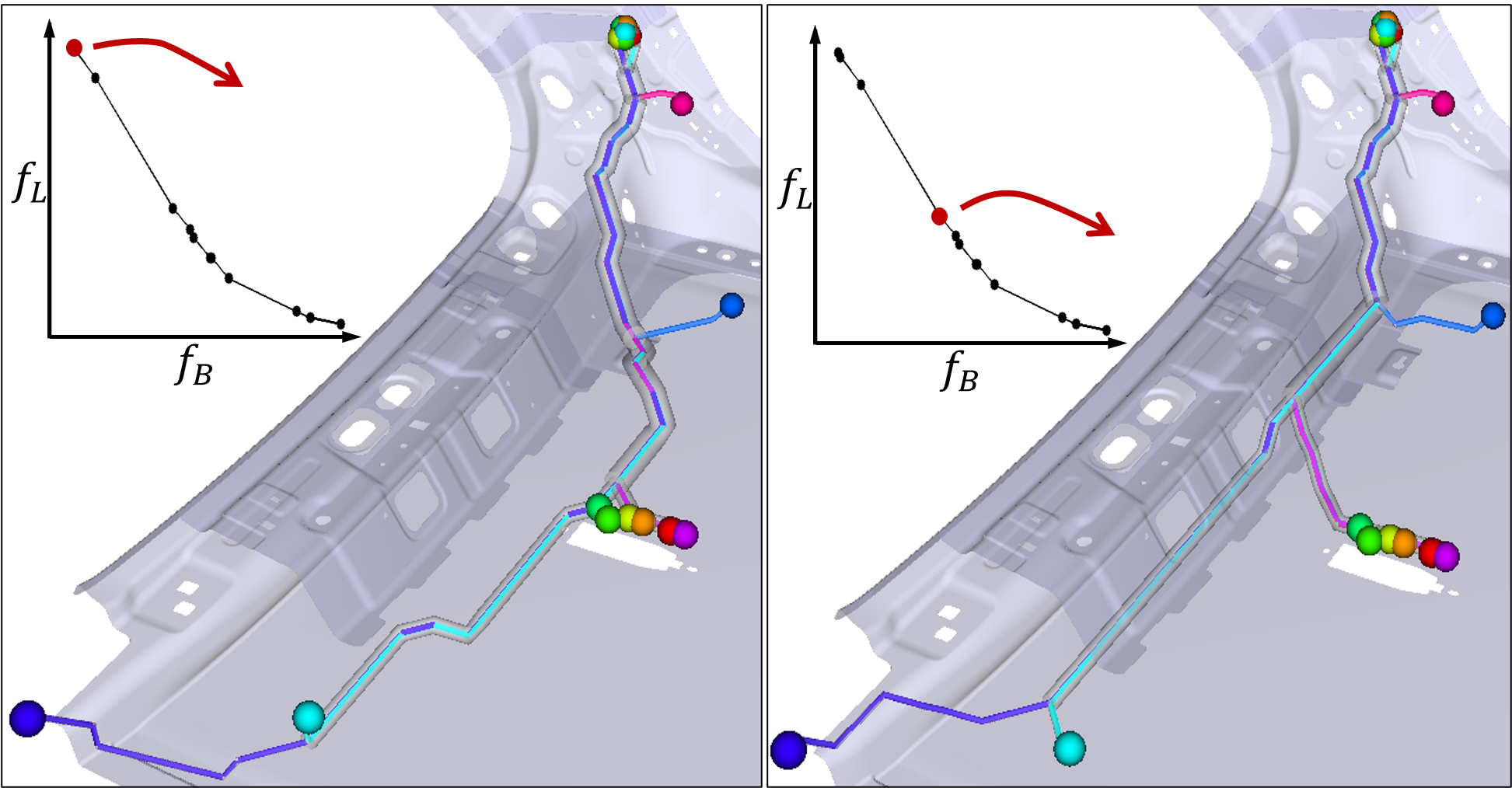}
		\caption{Case A.}
		\label{fig:case_A_example_solution}
	\end{subfigure}
	\begin{subfigure}[b]{0.45\textwidth}
		\centering
		\includegraphics[width=\textwidth]{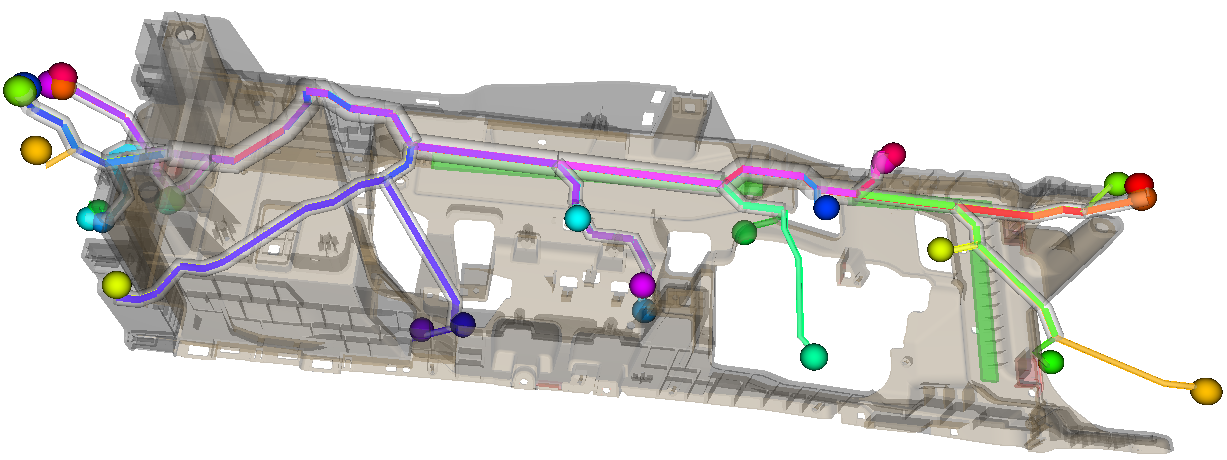}
		\caption{Case B.}
		\label{fig:case_B_example_solution}
	\end{subfigure}
	\begin{subfigure}[b]{0.45\textwidth}
		\centering
		\includegraphics[width=\textwidth]{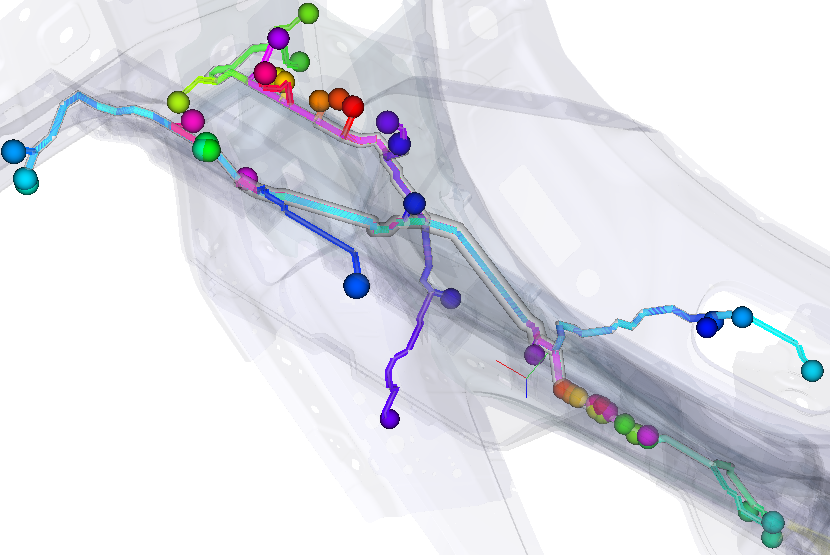}
		\caption{Case C.}
		\label{fig:case_C_example_solution}
	\end{subfigure}
	\caption{Example of some candidate solutions. The bundle radii are visualized as transparent grey geometry.}
	\label{fig:case_A-C_example_solution}
\end{figure}

Figure~\ref{fig:case_A_weight_vs_ub/lb} shows that we proved near-optimality for every value of $w_B$ for Case A (and likewise for Case B and Case C for some weights, see Figure~\ref{fig:case_B_lb_ub_vs_weights} and \ref{fig:case_C_lb_ub_vs_weights}). Example solutions for each case can be seen in Figure~\ref{fig:case_A-C_example_solution}. As discussed in Section \ref{sec:exact_solver}, the lower bound from the dual problem is not a reliable measure of optimality, as the optimal duality gaps can be large and therefore do not provide insight into the quality of our primal solution.

\begin{table}[!ht]
	\centering
	\caption{Computation time and number of unique solutions for the SHRH and $\alpha$-SPHRH.}
	\begin{tabular}{lrrr} 
		 & \multicolumn{1}{c}{Case A} & \multicolumn{1}{c}{Case B} & \multicolumn{1}{c}{Case C} \\ 
		\hline
		SHRH time & 19 s  & 222 s & 115 min \\ 
		SHRH \#sol. & 12 	& 91	& 446 \\
		\hline
		\rowcolor[gray]{0.9}
		$\alpha$-SPHRH time  & 7 s 	& 39 s	& 8 min \\
		\rowcolor[gray]{0.9}
		$\alpha$-SPHRH \#sol. & 9		& 19	& 49 \\
		\hline
	\end{tabular}
	\label{tab:industrial_cases_time_numsols}
\end{table}

\subsection{Design framework} \label{sec:design_framework}
Discussions with industrial professionals from Volvo Car Corporation have led to several design aspects and rules that should be considered in a cable harness design framework. Some of these items can be seen in Table~\ref{tab:design_rules} where the design process is categorized into three main steps and the check marks indicate which design aspects are suitable to handle at each step.

\newcommand{\cmarkg}{\textcolor{green!80!black}{\checkmark}}
\newcommand{\cmarky}{\textcolor{yellow!80!black}{\checkmark}}
\begin{table}[!ht]
	\centering
	\caption{Design aspects and rules to consider and indications in which algorithmic step they are suitable to handle. TR (Topology Router), SR (Smooth Router), S (Simulation). A green check mark means main impact and a yellow check mark means partial impact.}
	\begin{tabular}{ lccc } 
		\multicolumn{1}{c}{Design aspect} & TR & SR & S \\ 
		\hline
		Installation-friendly zones & \cmarkg & \cmarky & \\
		Cable-safe zones & \cmarkg	& \cmarky	& \\
		Clip-able geometry & \cmarkg & \cmarky & \\
		Clipping rules & \cmarky	& \cmarkg	& \\
		Terminal to branch point length & \cmarky	& \cmarkg	& \\
		Terminal directions & \cmarky	& \cmarkg	& \\
		Bundle radii &	& \cmarkg	& \\
		Minimum bend radii &	& \cmarkg	& \\
		Material physical properties &	& \cmarky	& \cmarkg \\
		\hline
	\end{tabular}
	\label{tab:design_rules}
\end{table}

\begin{figure}[!ht]
	\centering
	\begin{subfigure}[b]{0.45\textwidth}
		\centering
		\tcbox[size=fbox]{\includegraphics[width=\textwidth]{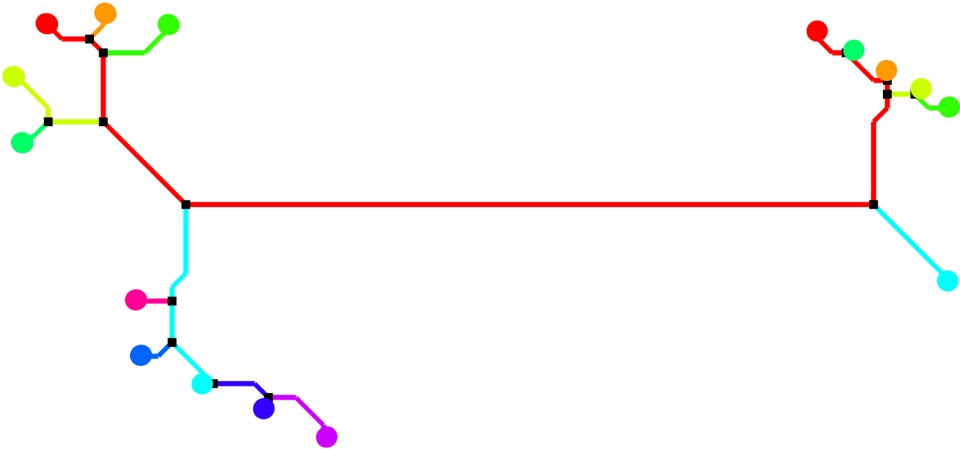}}
		\caption{}
		\label{fig:processing_pre}
	\end{subfigure}
	\begin{subfigure}[b]{0.45\textwidth}
		\centering
		\tcbox[size=fbox]{\includegraphics[width=\textwidth]{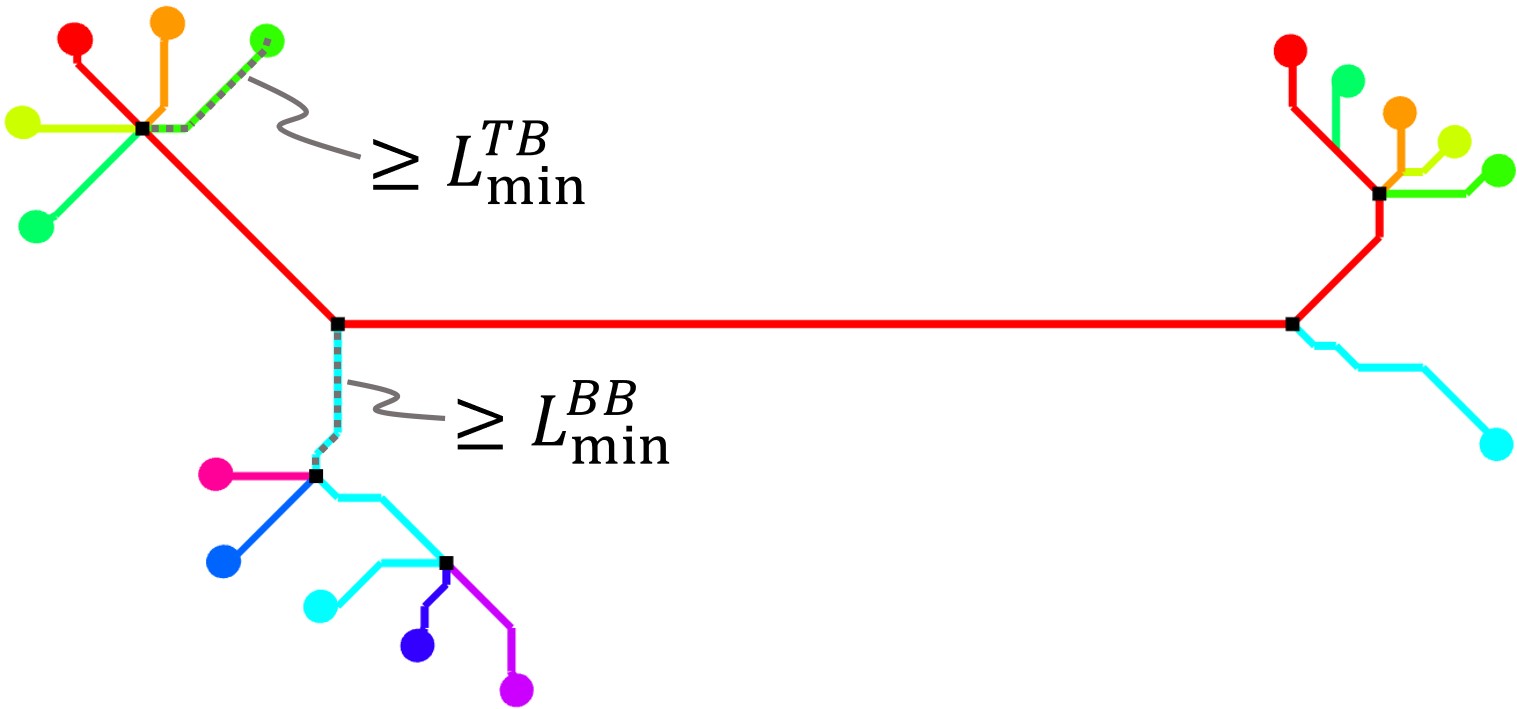}}
		\caption{}
		\label{fig:processing_post}
	\end{subfigure}
	\caption{(\subref{fig:processing_pre}) A solution to the CHRP. (\subref{fig:processing_post}) The solution after a local search for fulfilling constraints of minimum length between branch points ($L_{\text{min}}^{BB}$) and between a terminal and a branch point ($L_{\text{min}}^{TB}$). The black squares represent branch points.}
	\label{fig:processing}
\end{figure}

The contribution in this paper is part of the so-called \emph{topology router}. This router searches for solutions in the grid graph, determining the branch point locations and the main spatial location of the routes with respect to installation-friendly zones, cable-safe zones, and clip-able geometry. Clipping rules require at least one clip between two branch points and specify both minimum and maximum allowable distances between clips and between a clip and a branch point. We get minimum length constraints on the bundles due to the clipping rules and the requirement for a minimum length between a terminal and a branch point. Figure~\ref{fig:processing_post} depicts a solution that has been modified by a local search algorithm. This algorithm utilizes a greedy approach to either relocate or remove a branch point such that the minimum length constraints are fulfilled and $f$ is minimized. Moving a single branch point is done by summing the shortest distances from the connected branch points and terminals while constraint-violating nodes are excluded as potential branch points. It is important to consider the minimum length constraints in the topology router since the optimal number of branch points is affected by these rules.

The exact length of the bundles and the clip positions are decided by the so-called \emph{smooth router}---a tool currently in the prototype stage---that modifies the routes in the continuous 3D routing space with respect to the given topology such that the bundles are collision-free (with respect to their radii) and satisfy minimum bend radii (which depends on the cables' physical properties). The topology router incorporates terminal directions by penalizing edges behind terminals, this prevents clearly infeasible solutions. The exact terminal direction is used in the smoothing step, which is accelerated by the improved initial solution. 

As a final step, the cable harness should be simulated to static equilibrium with respect to its physical properties (e.g., stiffness) where deformation and tension due to gravity, contact, and clips can be analysed. 

\subsection{Comparison with related work} \label{sec:result_related_work}
In this section, we compare the objective function values and computation times obtained when using PSO (see \ref{appendix:pso}) and the $\alpha$-SPHRH. Also, a small 2D case from~\cite[Fig.~8]{masoudi2022optimization} was solved to note some differences and similarities between the CHRP and the model in~\cite{masoudi2022optimization}.

\begin{figure}[!ht]
	\centering
	\begin{subfigure}[c]{0.49\textwidth}
		\begin{overpic}[trim={1.9cm 0 1.75cm 0},clip,width=\textwidth]{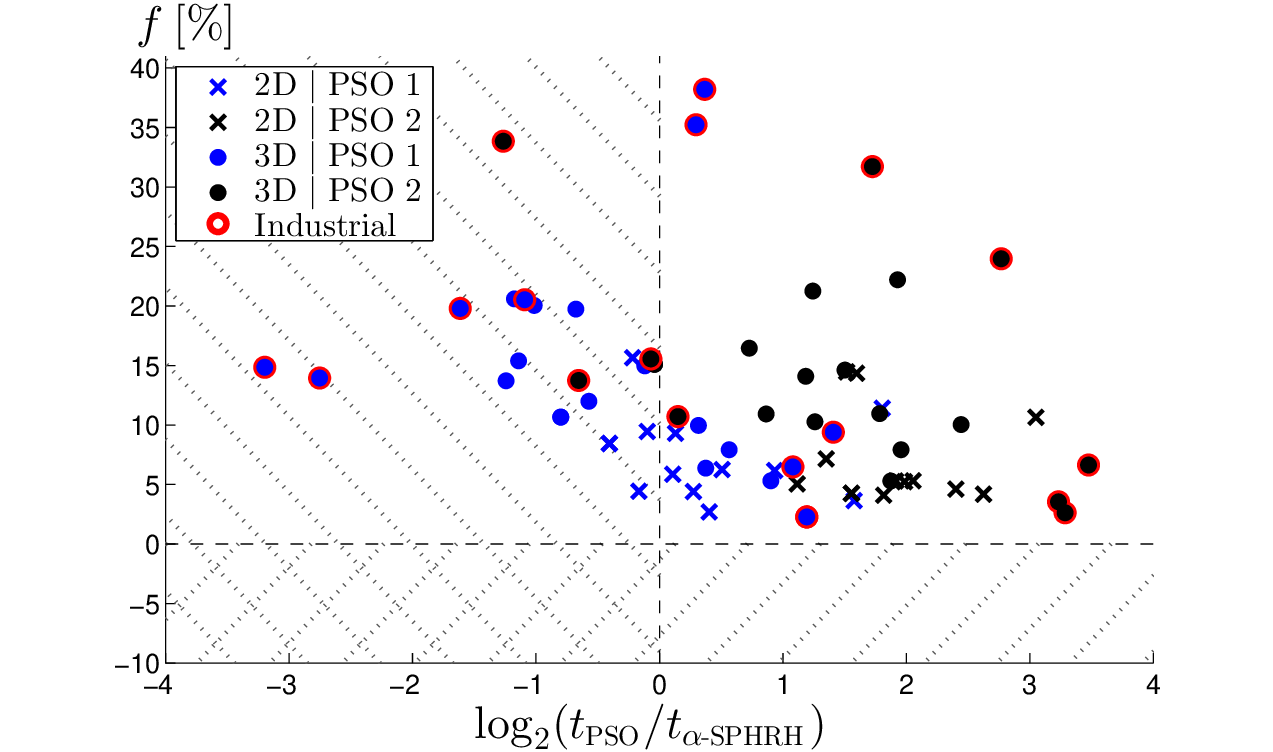}
			\put(65,12){\footnotesize \textbf{\emph{PSO better $f$}}}
			\put(8,21){\rotatebox{90}{\footnotesize \textbf{\emph{PSO better time}}}}
		\end{overpic}
	\end{subfigure}
	\caption{Relative difference in objective function value and computation time $t$ between PSO and $\alpha$-SPHRH. Each data point corresponds to a PSO setting, routing problem setup, and bundle weight.} 
	\label{fig:pso_vs_aSPHRH}
\end{figure}

The PSO method was compared to the $\alpha$-SPHRH for four 2D cases and seven 3D cases (including Case A--C). Each case was solved for $w_B\in\{0.1,\:0.3,\:0.6\}$ and two PSO parameter settings found in~\cite{zhang2021multi, zhang2021multi_PS}. The relative objective function value in Figure~\ref{fig:pso_vs_aSPHRH} was computed as $(f_{\text{PSO}} - f_{\alpha\text{-SPHRH}}) / f_{\alpha\text{-SPHRH}}$. The objective function evaluation was parallelized over the particles in the PSO implementation and the HRH was parallelized over the initial solutions in the $\alpha$-SPHRH. The results in Figure~\ref{fig:pso_vs_aSPHRH} indicate that the PSO approach, depending on the parameter setting, can be computationally comparable to the $\alpha$-SPHRH. It is worth noting that the absolute computation times range from 0.1 to 70 seconds, so although the absolute difference can be small, the relative time comparison can appear more extreme than it is for some instances. The big drawback with the PSO method is that it never found a better or equally good solution. The poor performance of the implemented PSO algorithm in minimizing $f$ indicates that the algorithm is not robust for the CHRP. A pattern search improved PSO algorithm was implemented according to~\cite{zhang2021multi_PS}, but the average improvement was only around 1\%--2\%. The computation time increased but is not presented due to non-optimized code implementation. Consequently, the originally suggested PSO algorithm serves as a better comparison. An advantage of the PSO method is that the objective function can easily be modified to represent other harness design factors. The stochastic behaviour is a drawback if solution reproducibility is desired, in contrast to the HRH which is deterministic, i.e., always produces the same solution for the same instance.

\begin{figure}[!ht]
	\centering
	\begin{subfigure}[c]{0.45\textwidth}
		\tcbox[size=fbox]{\begin{overpic}[width=\textwidth]{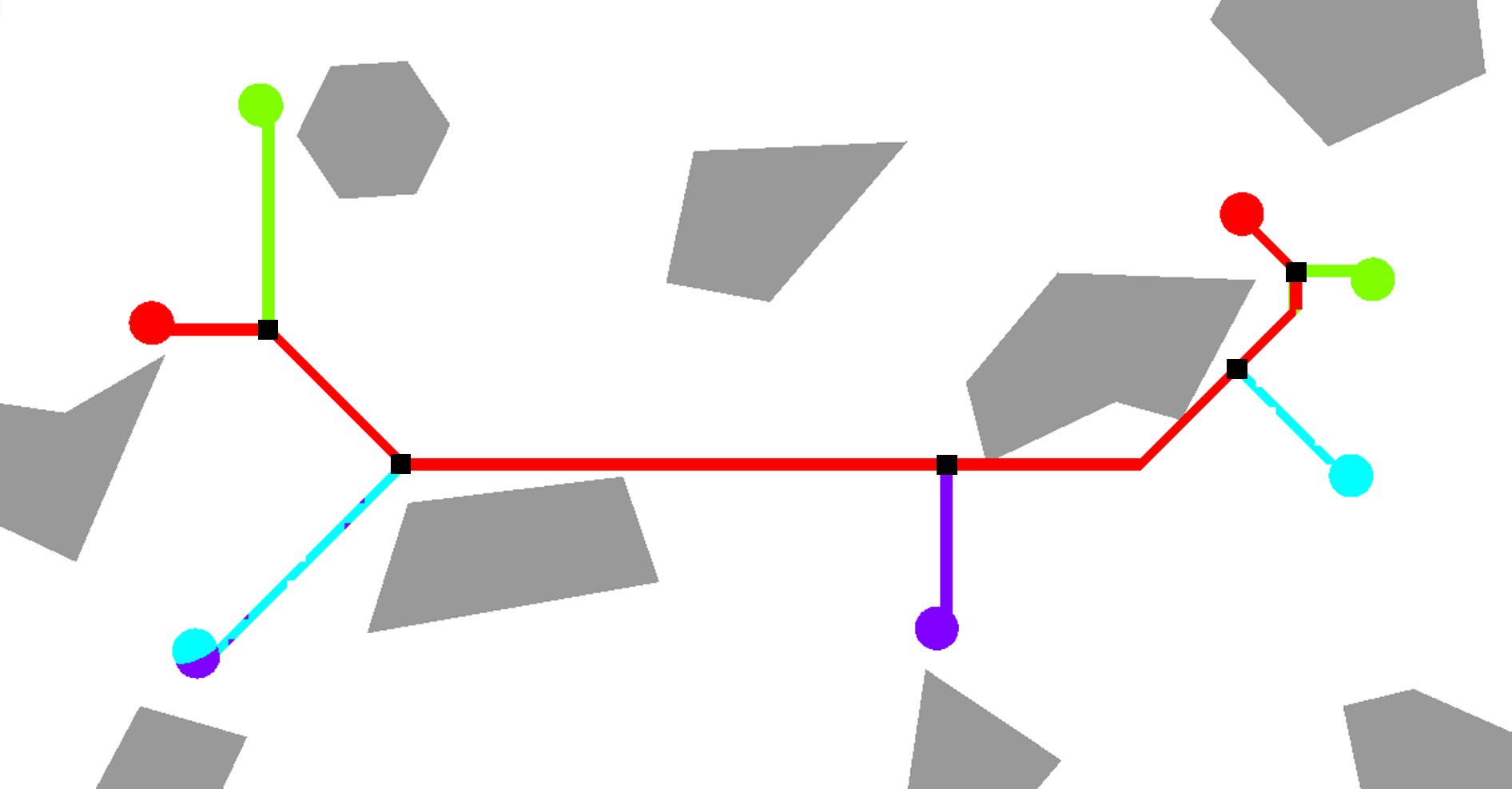}
				\put(20,41){
					\tcbox[size=fbox]{
						\begin{minipage}{13.5em}
							\small Total cable length = 239 cm \\
							Common length = 55 cm
						\end{minipage}
					}
				}
		\end{overpic}}
	\end{subfigure}
	\caption{Example solution from solving the instance in~\cite[Fig.~8]{masoudi2022optimization} with the $\alpha$-SPHRH parallelized over 15 bundle weights (computation time: 0.2 seconds). The common length is the sum of lengths between all branch points (black squares).}
	\label{fig:masoudi_compare_scene_aphh_solutions}
\end{figure}

Objective functions (\ref{eq:obj_functions}) and the ones in~\cite{masoudi2022optimization} are formulated differently, but both formulations correspond to minimizing the total cable lengths and favour common paths. The main difference between the models is that the number of branch points is specified in~\cite{masoudi2022optimization}, whereas in our paper, it is an outcome of optimizing the CHRP. The harness topology in Figure~\ref{fig:masoudi_compare_scene_aphh_solutions} resulted in five branch points and has about the same total length but more common length compared to Cable Harness Layout \#7 in~\cite[Fig.~7]{masoudi2022optimization} which has two branch points. The optimal number of branch points depends on the specific case, and determining what is optimal may lack a straightforward measurement. It could be influenced by subjective preferences of the designer or guided by design rules as depicted in Figure~\ref{fig:processing_post}.

\section{Conclusion and future work} \label{sec:conclusion_and_future_work}
This paper has presented a novel method to find a variety of cable harness topologies in a 3D environment that are optimized with respect to the minimum length of the cables, rewarding common paths, and the spatial location of the routes. The spatial location depends on case-specific factors which can be preference inputs from the designer, e.g., avoiding hazardous zones. An exact solver is not computationally affordable, and a deterministic local search method has been developed that can provide solutions close to the Pareto front efficiently (in magnitude of seconds to minutes for industrial-sized cases). Because the proposed model does not capture all design aspects, it is important to find a variety of promising candidate solutions that will be further processed. Topologically different solutions were found by optimizing for a set of objective function weights and by locally optimizing different initial routings.

\paragraph{Design framework}
One significant question for future work is how the developed method can be used in a framework that guides the designer throughout the whole process from setting up the routing environment to a final simulated harness design that can be evaluated in a virtual reality environment. The idea is that the presented method is the first step in this framework where the next steps use a set of the generated harness topology candidates and optimize over more design factors, such as bend radii and clip positions.

\paragraph{Field cost customization}
We have shown for some simple cases that case-specific preference input can be given from the designer to yield certain routing designs. Further investigation is needed to determine the design factors suitable to be represented through the cost field. It is also important to ascertain whether human input is necessary, or if the design factors can be encoded automatically. For example, capturing both feasible and ergonomic aspects of harness installation in this early stage is beneficial. These aspects can be considered by utilizing a digital human modelling simulation software tool, e.g., IMMA~\cite{delfs2014automatic} and Jack~\cite{blanchonette2010jack}, which enables us to evaluate if certain zones are reachable and ergonomic for a human and adjust the cost field accordingly.

\section*{Acknowledgments}
We thank our research partners at the University of Linköping and Volvo Car Corporation. We also thank Dr Masoudi at Clemson University for providing the problem instance in Figure~\ref{fig:masoudi_compare_scene_aphh_solutions} used for comparison in Section \ref{sec:result_related_work}. This work was carried out within the Production in a circular economy Area of Advance at Chalmers University of Technology and the research project AutoPack 2.0 (DNR 2020-05173) funded by Sweden’s innovation agency Vinnova.

\section*{Declaration of Generative AI and AI-assisted technologies in the writing process}
During the preparation of this work the authors used ChatGPT in order to improve language and grammar. After using this tool, the authors reviewed and edited the content as needed and take full responsibility for the content of the publication.

\appendix 
\section{PSO details}\label{appendix:pso}
\begin{figure}[!ht]
	\centering
	\begin{subfigure}[t]{0.5\textwidth}
		\centering
		\includegraphics[trim={0 0.27cm 0 0},clip,width=\textwidth]{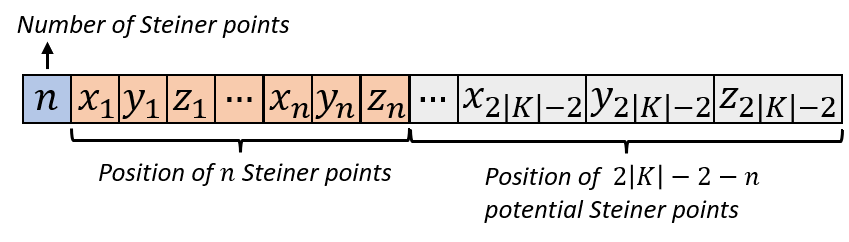}
		\caption{Particle encoding. The maximum number of Steiner points is $2\vert K\vert-2$.}
		\label{subfig:PSO_solution_particle_encoding}	
	\end{subfigure}
	\vspace{5pt}
	\begin{subfigure}[t]{0.23\textwidth}
		\centering
		\includegraphics[width=\textwidth]{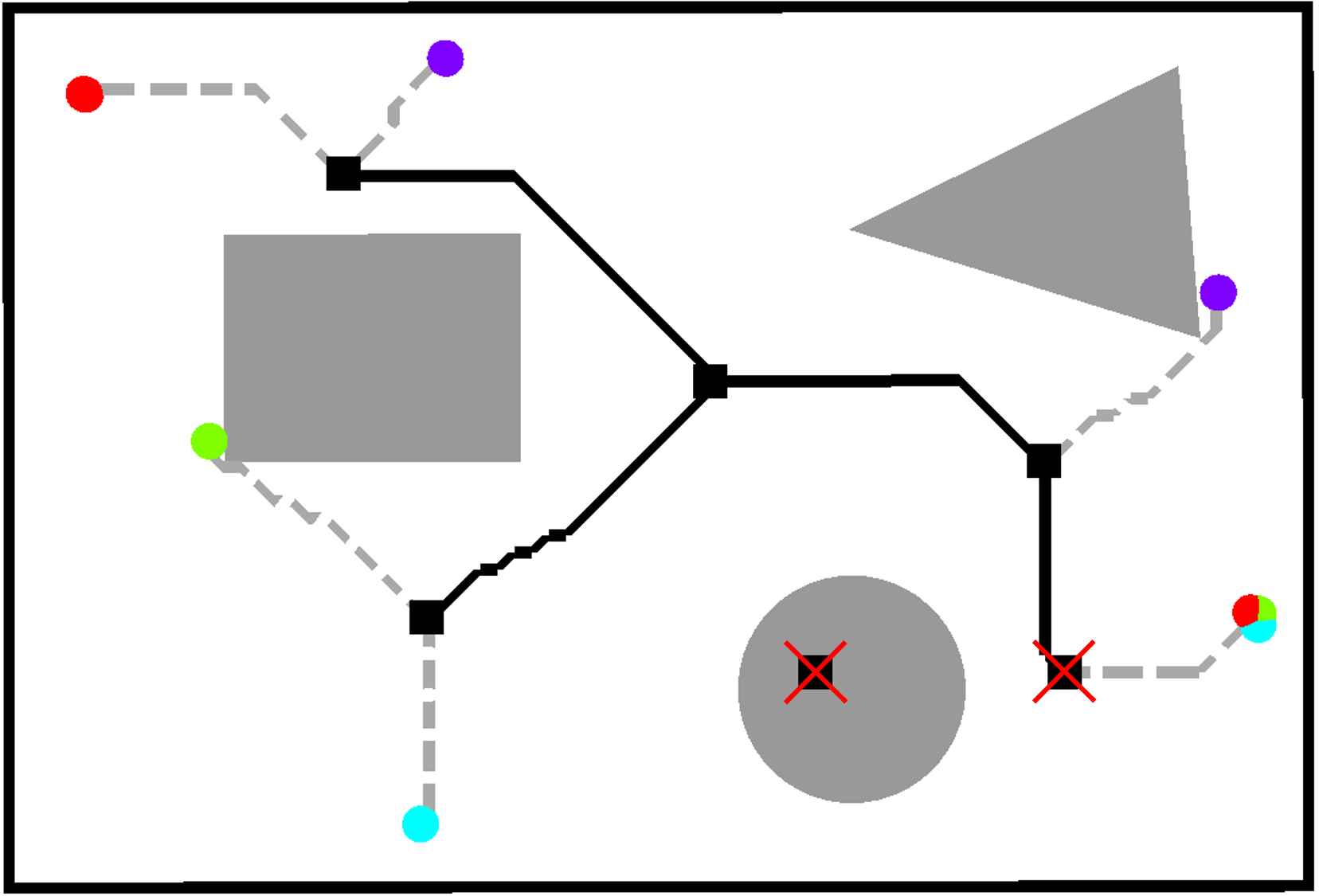}
		\caption{Decoded particle where the Steiner points correspond to the black squares. Unreasonable Steiner points are removed (crossed out).}
		\label{subfig:PSO_solution_MST}	
	\end{subfigure}
	\hspace{1mm}
	\begin{subfigure}[t]{0.23\textwidth}
		\centering
		\includegraphics[width=\textwidth]{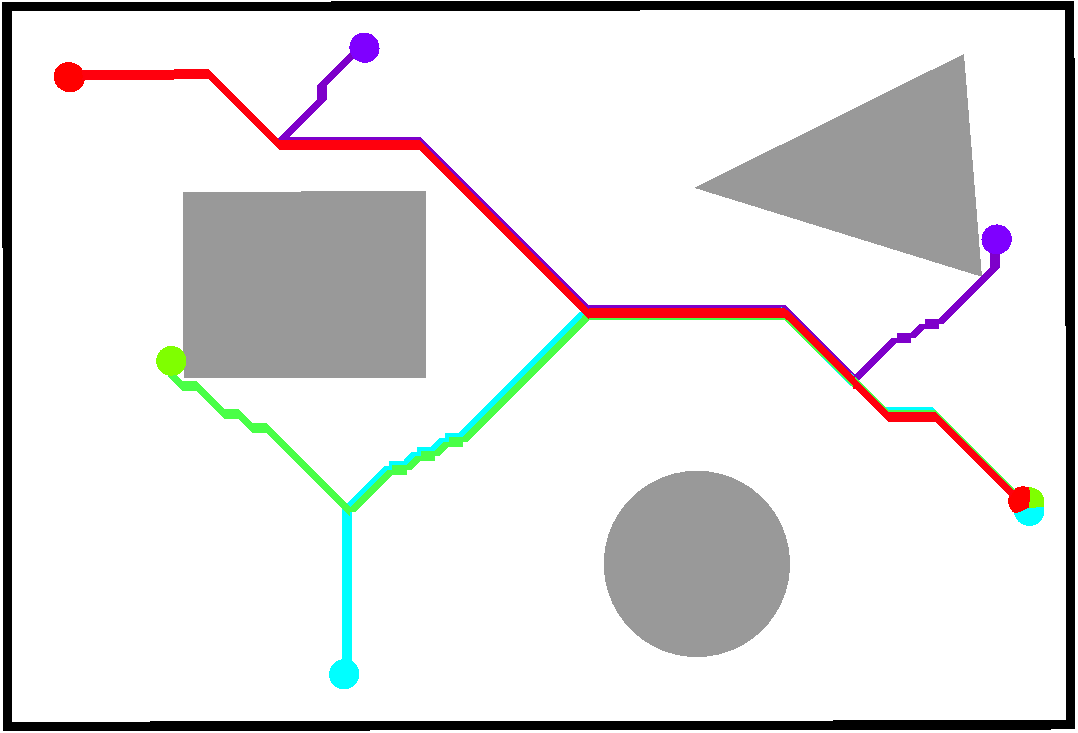}
		\caption{Resulting harness solution.}
		\label{subfig:PSO_solution_harness}
	\end{subfigure}
	\caption{Construction of a harness topology from a PSO-particle.}
	\label{fig:PSO_solution}
\end{figure}

A \emph{Particle Swarm Optimization (PSO)} algorithm for finding cable harness topologies is used for comparison in Section \ref{sec:result_related_work}. The algorithm is based on the methods described in~\cite{zhang2021multi, zhang2021multi_PS, liu2012multi} which all use the particle encoding depicted in Figure~\ref{subfig:PSO_solution_particle_encoding}. The PSO algorithm is implemented according to the pseudocode in~\cite{liu2012multi}. The particle encoding has a fixed length equal to $3(2\vert K\vert-2)+1$ and contains the number of $n$ Steiner points (branch points), the position of these $n$ points, and the position of $2\vert K\vert-2-n$ potential Steiner points. The position of a Steiner point is mapped to the closest grid point in our case and Steiner points in collision are disregarded.

To evaluate a particle a \emph{Minimum Spanning Tree (MST)} needs to be constructed based on a graph consisting of edges between all Steiner points and edges between the Steiner points and terminals. These edges correspond to the shortest paths. The MST is found efficiently by using a modified Dijkstra's algorithm: a tree is initialized with a single Steiner point, and the shortest paths to all nodes from the Steiner points in the tree are searched. A Steiner point or terminal is added and connected to the tree when they are visited. The search is terminated when all Steiner points and terminals have been connected. An MST given some Steiner points can be seen in \ref{subfig:PSO_solution_MST}, where points that are connected to fewer than three other nodes in the MST are removed, as in~\cite{zhang2021multi_PS}, resulting in the final solution in Figure~\ref{subfig:PSO_solution_harness}.

\bibliography{manuscript}
\end{document}